\documentclass[10pt,twocolumn]{article}

\usepackage[margin=0.75in,columnsep=0.25in]{geometry}
\usepackage[T1]{fontenc}
\usepackage[utf8]{inputenc}
\usepackage{mathptmx}
\usepackage{graphicx}
\usepackage{booktabs}
\usepackage{array}
\usepackage{ragged2e}
\usepackage{microtype}
\usepackage{caption}
\usepackage[hidelinks]{hyperref}
\usepackage{xurl}
\usepackage{titlesec}
\usepackage{balance}
\usepackage{placeins}
\titlespacing*{\section}{0pt}{2.0ex plus .6ex minus .2ex}{1.0ex plus .2ex}
\titlespacing*{\subsection}{0pt}{1.6ex plus .5ex minus .2ex}{0.8ex plus .2ex}
\titlespacing*{\subsubsection}{0pt}{1.4ex plus .4ex minus .2ex}{0.6ex plus .2ex}
\titleformat*{\section}{\large\bfseries}
\titleformat*{\subsection}{\normalsize\bfseries}
\titleformat*{\subsubsection}{\normalsize\bfseries\itshape}

\newcommand{\ul}[1]{#1}
\newcommand{\hl}[1]{#1}
\newcommand{\orc}[2]{\href{https://orcid.org/#2}{#1}}

\begin{document}

\twocolumn[{%
\centering
{\LARGE\bfseries Rethinking Generative AI Literacy: An Integrative, Developmental,\\[2pt]
and Dialectical Framework for K--12 Teacher Education\par}
\vspace{1.2em}
{\large
\orc{Shahin Hossain}{0000-0002-3461-1147}\textsuperscript{1}, \orc{Sima Ahmadi}{0009-0007-6637-8638}\textsuperscript{2}, \orc{Leqi Li}{0009-0004-3232-1949}\textsuperscript{3}, \orc{Idowu David Awoyemi}{0009-0001-3112-7712}\textsuperscript{4}, \orc{Wei Huang}{0009-0006-2453-9715}\textsuperscript{5},\\[2pt]
\orc{Chenxi Zhou}{0009-0002-4545-6274}\textsuperscript{6}, \orc{Jujia Li}{0009-0001-7982-5122}\textsuperscript{7}, \orc{Samaa Haniya}{0000-0001-9676-2242}\textsuperscript{8}, \orc{Shapla Khanam}{0000-0003-4012-086X}\textsuperscript{9}, Tasbirun Mashreka Subaha\textsuperscript{10}
\par}
\vspace{0.9em}
{\small
\textsuperscript{1}School of Education, University of Maryland Baltimore County, Baltimore, Maryland, USA \,\textbullet\, \href{mailto:shahinh1@umbc.edu}{shahinh1@umbc.edu}\\
\textsuperscript{2}School of Teaching, Learning, and Curriculum Studies: Educational Technology Department, Kent State University, Ohio, USA \,\textbullet\, \href{mailto:sahmadi@kent.edu}{sahmadi@kent.edu}\\
\textsuperscript{3}Department of Learning and Performance Systems, Pennsylvania State University, State College, PA, USA \,\textbullet\, \href{mailto:lzl5599@psu.edu}{lzl5599@psu.edu}\\
\textsuperscript{4}Department of Educational Leadership, Policy, and Technology Studies, The University of Alabama, Tuscaloosa, AL, USA \,\textbullet\, \href{mailto:idawoyemi@crimson.ua.edu}{idawoyemi@crimson.ua.edu}\\
\textsuperscript{5}College of Education, University of Alabama, Alabama, USA \,\textbullet\, \href{mailto:whuang20@crimson.ua.edu}{whuang20@crimson.ua.edu}\\
\textsuperscript{6}Graduate School of Education and Human Development, Department of Curriculum and Instruction, George Washington University, District of Columbia, USA \,\textbullet\, \href{mailto:chenxi.zhou1@gwu.edu}{chenxi.zhou1@gwu.edu}\\
\textsuperscript{7}College of Education, University of Alabama, Alabama, USA \,\textbullet\, \href{mailto:jli183@crimson.ua.edu}{jli183@crimson.ua.edu}\\
\textsuperscript{8}Graduate School of Education and Psychology, Pepperdine University, Los Angeles, CA, United States \,\textbullet\, \href{mailto:samaa.haniya@pepperdine.edu}{samaa.haniya@pepperdine.edu}\\
\textsuperscript{9}Department of Artificial Intelligence, Faculty of Computer Science and Information Technology, Universiti Malaya, Kuala Lumpur 50603, Malaysia \,\textbullet\, \href{mailto:shapla.k@um.edu.my}{shapla.k@um.edu.my}\\
\textsuperscript{10}Department of English Language and Literature, Begum Rokeya University, Rangpur, Bangladesh \,\textbullet\, \href{mailto:tasbirunmashreka.12102020@student.brur.ac.bd}{tasbirunmashreka.12102020@student.brur.ac.bd}
\par}
\vspace{1.6em}
}]

\section*{Abstract}
Generative artificial intelligence (GenAI) has entered classrooms faster
than teachers have been prepared to use it well, producing a GenAI
literacy lag in which technological diffusion outpaces
educators\textquotesingle{} conceptual, pedagogical, and ethical
readiness. Established AI literacy frameworks predate the widespread
adoption of large language models and, while acknowledging ethics,
position it as a discrete competency rather than a constitutive
commitment, with equity and agency treated as supplementary design
principles. More recent GenAI-specific efforts address isolated features
but remain fragmented. This paper introduces the Responsible AI Literacy
in Education (RAIL-Ed) framework, developed through a systematic review
and qualitative framework analysis of 67 studies (2023--2025) coded
against five leading frameworks and grounded in critical, pragmatist,
sociocultural, and human-centered traditions (Freire, Dewey, Vygotsky,
and Shneiderman). RAIL-Ed specifies six interdependent pillars:
Technical Fluency, Critical Evaluation, Human--AI Collaboration,
Contextual Awareness, Ethical Reasoning, and Empowered Agency,
distinguished by three commitments. It is integrative: the absence of
any pillar produces a characteristic pedagogical failure. It is
developmental: a three-level rubric (Emerging, Competent, Advanced)
specifies how each pillar matures across the K--12 teacher-preparation
continuum. And it is dialectical: the same generative affordance can
deepen or displace learning depending on the literacy a teacher brings
to it, making the cultivation of that literacy, not the adoption of the
tool, the object of design. By treating ethics, equity, and agency as
constitutive, RAIL-Ed offers a theoretically grounded basis for
curriculum design, teacher education, and policy, aligned with the
UNESCO AI Competency Framework for Teachers and the OECD/European Commission AILit Framework. The framework is conceptual, advancing
falsifiable propositions for empirical validation.

\medskip
\noindent\textbf{Keywords:} Generative AI literacy; teacher education; critical AI literacy; human--AI collaboration; AI ethics in education; teacher professional development
\vspace{0.6em}

\section{Introduction}\label{introduction}

The release of ChatGPT in late 2022 and the rapid proliferation of
generative artificial intelligence (GenAI) systems that followed mark
one of the most consequential shifts in educational technology in recent
history (Kasneci et al., 2023; Miao \& Holmes, 2023). Unlike earlier
technologies such as Web 2.0, the Internet of Things, learning
management systems, or rule-based intelligent tutors, generative systems
such as ChatGPT, Claude, Gemini, and Copilot differ qualitatively in
three ways: they generate novel content rather than retrieve it,
interact through open-ended natural language rather than predetermined
rules, and produce fluent outputs whose accuracy and provenance resist
verification. They generate content in real time, such as essays, code,
images, audio, problem solutions, and arguments that learners and
teachers increasingly treat as collaborative outputs (Mollick \&
Mollick, 2023). This shift unsettles long-standing educational
assumptions about authorship, expertise, and cognition, and invites
consideration of how generative systems function as epistemic mediators
in learning contexts (Bender et al., 2021; Kay et al., 2024).

Educational institutions therefore confront a layered responsibility.
They must prepare learners to engage productively with generative
systems while simultaneously cultivating the critical, ethical, and
reflective sensibilities required to interrogate them (Selwyn, 2022;
Miao \& Holmes, 2023). International policy standards have begun to
articulate this responsibility through frameworks that emphasize human
oversight, equity, and transparency (OECD, 2024; OECD/European
Commission, 2026; Miao \& Holmes, 2023). Translating these high-level
commitments into coherent classroom practice remains unresolved, in part
because GenAI tools evolve faster than curricula, teacher preparation
programs, and institutional policy (Kasneci et al., 2023; Miao \&
Holmes, 2023).

A growing body of scholarship documents what may be characterized as a
GenAI literacy lag: the widening gap between the speed at which GenAI
tools enter classrooms and the slower development of the conceptual,
pedagogical, and ethical capabilities educators and learners need to use
them well (Cheah et al., 2025; Sattelmaier \& Pawlowski, 2025; Sperling
et al., 2024). The result is that many users can operate these systems
without being equipped to evaluate their outputs, recognize their
limitations, or apply them in pedagogically sound ways. Existing AI
literacy frameworks, although foundational, were developed largely in
response to earlier paradigms of machine learning and rule-based
classification (Long \& Magerko, 2020; Ng et al., 2021). They tend to
emphasize technical comprehension and instrumental skill over
pedagogical principle, operationalizing literacy through relatively
stable, classifier-style applications that presuppose standardized
baselines and competency checklists.

Long and Magerko\textquotesingle s (2020) framework, for instance,
specifies seventeen discrete competencies, while Ng et al. (2021)
suggested four dimensions of AI literacy (know and understand AI, use
and apply AI, evaluate and create AI, and AI ethics) for students. Such
models emerged largely in response to predictive, classifier-based
systems, whose bounded outputs let learners form stable mental models of
how a system behaves. Generative systems introduce different demands:
they depend heavily on prompt construction and rhetorical framing
(Mollick \& Mollick, 2023); they produce probabilistic and frequently
inaccurate outputs commonly described as hallucinations (Bender et al.,
2021; Kasneci et al., 2023); and they encode representational biases at
scale (Lee et al., 2024). They can also invite forms of cognitive
offloading that erode learning when used without scaffolding (Kasneci et
al., 2023; Selwyn, 2022).

Recent frameworks increasingly acknowledge these tensions individually,
with growing attention to ethical reasoning, prompt literacy, and
critical evaluation of outputs (Jin et al., 2025; Ng et al., 2025;
Ofosu-Asare, 2025). Although epistemic harms have been theorized in
unified terms outside education (Kay et al., 2024), research rarely
integrates hallucination, prompt dependency, epistemic risk, and
AI-mediated inequity into a single theoretical framework for teaching
and learning (Park, 2025). The absence of an integrative framework that
connects technical fluency with critical, ethical, and equity-oriented
capacities leaves teachers underprepared for the responsibilities GenAI
imposes on practice (Lee et al., 2024; Miao \& Holmes, 2023).

This paper introduces the Responsible AI Literacy in Education (RAIL-Ed)
framework to address this gap. We synthesized existing scholarship
through a systematic review and qualitative framework analysis of the AI
literacy literature. The framework\textquotesingle s epistemological
foundation draws on four traditions: three educational traditions and
one human-centered design tradition. The educational traditions are
Freire\textquotesingle s (1970/2000) pedagogy of critical consciousness,
Dewey\textquotesingle s (1938) reflective inquiry, and
Vygotsky\textquotesingle s (1978) sociocultural theory of mediated
cognition. The fourth, Shneiderman\textquotesingle s (2020)
human-centered AI principles, which pair strong human control and
oversight with high levels of automation, positions the framework within
responsible system design. From this synthesis, RAIL-Ed articulates six
interdependent pillars: Technical Fluency, Critical Evaluation,
Human--AI Collaboration, Contextual Awareness, Ethical Reasoning, and
Empowered Agency. What distinguishes RAIL-Ed is not its pillars alone
but three commitments that govern how they operate together: it is
integrative, developmental, and dialectical. This framework is intended
for K--12 teacher education and is adaptable to diverse disciplinary,
cultural, and institutional contexts (Ng et al., 2021; Miao \& Holmes,
2023). It addresses the entire K--12 teacher-preparation continuum,
encompassing both pre-service teachers in undergraduate programs and
in-service teachers currently in classrooms, where the literature
identifies the most significant gap in preparation. These groups are
treated as developmentally connected points along a single continuum.
Teachers are positioned at the center of this framework because
teachers' GenAI literacy influences the literacy of every cohort of
students they subsequently instruct, making teacher preparation the most
impactful site for intervention. The scope of this work is conceptual;
it does not provide empirical validation of the framework or present it
as a finalized product. Instead, it offers a theoretically grounded
scaffold intended to inform curriculum design, teacher education,
professional development, and policy (OECD, 2024; Tan, 2025).

This paper makes three primary contributions. First, it argues that
responsible engagement with GenAI requires pedagogically and
theoretically grounded foundations, framing substantive learning as a
deliberate design commitment. Second, it treats ethics, equity, and
agency as constitutive elements of literacy, responding to concerns
about representational harm (Bender et al., 2021; Lee et al., 2024),
epistemic injustice (Kay et al., 2024), and structural inequity in
AI-mediated learning (Selwyn, 2022). Third, it offers a framework whose
dimensions are interdependent and mutually constitutive, conceptualizing
literacy as a relational, situated, and ethical practice. Through these
contributions, the paper advances scholarly and policy discussions on
what responsible GenAI integration meaningfully requires of teachers,
learners, and the broader educational community (OECD, 2024; Tan, 2025).

The remainder of the paper proceeds as follows. Section 2 establishes
RAIL-Ed\textquotesingle s theoretical foundations and reviews existing
AI and GenAI literacy frameworks; Section 3 identifies the gaps that
motivate a new framework; Section 4 presents RAIL-Ed, detailing its six
pillars, failure-mode map, and developmental rubric; Section 5 develops
implications for practice, policy, and research; and Sections 6 and 7
address limitations, future directions, and conclusions.

\section{Theoretical Foundations: Critical, Pragmatist,
Sociocultural, and Human-Centered
Traditions}\label{theoretical-foundations-critical-pragmatist-sociocultural-and-human-centered-traditions}

A framework for GenAI literacy requires more than an inventory of
competencies; it requires a theory of how teachers and learners come to
know, judge, and act with a technology that mediates knowledge itself.
RAIL-Ed draws on four complementary traditions to supply that theory.
Freire's critical pedagogy establishes literacy as an exercise of
agency. Dewey's pragmatism establishes it as a reflective inquiry.
Vygotsky's sociocultural psychology establishes cognition as mediated by
cultural tools; and Shneiderman's human-centered design establishes the
conditions under which automation augments rather than displaces human
judgment. These traditions reject the deficit thinking that has
historically framed learners' difficulties as personal deficiencies, and
they position equity not as a topic to be added but as a property
constitutive of literacy itself. The subsections that follow develop
each tradition and link it to the pillars it grounds.

\subsection{A Critical Foundation: Rejecting Deficit Thinking
(Freire)}\label{a-critical-foundation-rejecting-deficit-thinking-freire}

RAIL-Ed begins from a critical commitment: a rejection of the deficit
thinking that attributes students' academic difficulties to perceived
deficiencies in learners, families, or communities (Davis \& Museus,
2019). Such framing, prominent across twentieth-century educational
discourse, justified lowered expectations and remedial tracking for
students from marginalized backgrounds (Tewell, 2020). It persists today
whenever learners' struggles with new technologies are read as
individual readiness gaps. Freire's (1970/2000) critique of the
``banking model'' in which teachers deposit information into passive
students names the pedagogy that deficit thinking produces. In
opposition to it, Freire proposed a problem-posing education grounded in
dialogue, reflection, and praxis, in which learners are active
participants capable of transforming their conditions through critical
inquiry. This commitment is directly consequential for GenAI literacy. A
deficit framing reduces literacy to operational compliance, prompt
construction, and rule-following while obscuring how AI systems shape
knowledge, authority, and power (Selwyn, 2022). Freire's tradition
instead requires that teachers be prepared to help students interrogate
those systems, not merely operate them. It grounds three RAIL-Ed
pillars: Contextual Awareness, Ethical Reasoning, and Empowered Agency,
positioning equity and ethics as foundational to AI literacy.

\subsection{A Pragmatist Foundation: Reflective Inquiry (Dewey)}

If the critical tradition establishes why literacy must attend to power,
the pragmatist tradition establishes how it is learned: through
experience, experimentation, and reflection. Dewey (1938) understood
learning as a social and democratic process in which knowledge is
constructed through active inquiry and the disciplined examination of
one's own experience. Applied to GenAI, Dewey's stance reframes the
central pedagogical question from whether students can use a tool to
what intellectual work the tool performed, displaced, or privileged in a
given encounter, and what that means for the learner's developing
judgment. Reflection, in this view, is not an end-of-task formality but
a structural feature of practice. Grounded in critical and pragmatist
traditions, RAIL-Ed\textquotesingle s Critical Evaluation pillar
combines Freire\textquotesingle s emphasis on questioning AI authority
with Dewey\textquotesingle s emphasis on reflective inquiry.

\subsection{A Sociocultural Foundation: Mediated Cognition (Vygotsky)}

While the critical and pragmatist traditions concern the purposes and
processes of learning, Vygotsky's (1978) sociocultural psychology
concerns its mechanisms. For Vygotsky, higher mental functions develop
through the mediation of cultural tools, language, symbol systems, and
technologies that not only assist thought but also reshape it. Cognition
is therefore never wholly individual; it is distributed across the
learner, more capable others, and the tools through which knowledge is
constructed. Generative systems represent a mediating tool of unusual
power: unlike a calculator or a search engine, a large language model
(LLM) participates in the formation of arguments, interpretations, and
explanations, and recent scholarship has begun to theorize it explicitly
as a mediational agent within sociocultural learning (Tate et al.,
2026). This grounding carries a sharp implication that the framework
operationalizes: a teacher who does not understand how the tool mediates
the production, bias, and bounds of its outputs is mediated by it.
Vygotsky\textquotesingle s tradition thus grounds
RAIL-Ed\textquotesingle s Technical Fluency pillar, which treats
functional understanding of the model as a prerequisite for control, and
its Human--AI Collaboration pillar, which positions the system as a
mediational agent in teaching, participating in the work without
occupying the social position of a partner (Tate et al., 2026). It also
informs Contextual Awareness, since what a tool mediates depends on the
cultural and linguistic setting in which it is used.

\subsection{A Human-Centered Design Foundation
(Shneiderman)}\label{a-human-centered-design-foundation-shneiderman}

The three educational traditions explain how literacy is proposed,
learned, and mediated; a fourth specifies the design principle that
should govern the human--AI relationship itself. Shneiderman's (2020)
human-centered AI rejects the common assumption that human control and
machine automation trade off on a single continuum. He argues instead
that they are independent dimensions, and that the most reliable, safe,
and trustworthy systems occupy the quadrant of high automation and high
human control at once. This reframing matters for education because it
dissolves the false choice between banning GenAI and surrendering
pedagogical judgment to it: the goal is neither minimal automation nor
minimal oversight but high levels of both, with the teacher retaining
control over consequential decisions. The principle aligns with the
human-in-the-loop tradition in AI-in-education research, in which human
judgment retains oversight of AI-mediated processes (Memarian \& Doleck,
2024), while extending it from system design to pedagogical practice.
Shneiderman's tradition grounds RAIL-Ed's Human--AI Collaboration
pillar, which asks teachers to maximize the system's generativity
without ceding authority over it, and its Empowered Agency pillar, in
which sustained human control is exercised through principled adoption,
restraint, or refusal. Shneiderman himself resists describing AI as a
teammate or partner, holding that computers should support human
activity. RAIL-Ed combines Vygotskian mediation with
Shneiderman\textquotesingle s control--automation principle to support
collaboration while maintaining pedagogical authority.

\subsection{Synthesis: From Four Traditions to One
Lens}\label{synthesis-from-four-traditions-to-one-lens}

These four traditions are not invoked as parallel citations but as a
single, integrated lens. The critical tradition supplies the framework's
purpose (literacy as agency and equity), the pragmatist tradition its
method (reflective inquiry), the sociocultural tradition its mechanism
(mediated cognition), and the human-centered tradition its design
principle (control and automation maximized together). Their synthesis
yields the framework's central and distinctive claim. Because cognition
is mediated by tools (Vygotsky) but those tools operate within
structures of power and possibility (Freire), the same generative
affordance does not carry a fixed value: it can deepen inquiry or
displace it, widen access or entrench advantage, depending on the
literacy the teacher brings to the encounter. RAIL-Ed names this its
dialectical commitment, and it follows from the theory. Equity cannot be
treated as a separate competency because literate AI practice requires
attention to the structural conditions shaping how tools are accessed
and used.

The pillars developed in Section 4 operationalize this lens; the gaps
that necessitate such a framework are documented first in Section 3.

\begin{figure*}[t]
\centering
\includegraphics[width=0.88\textwidth]{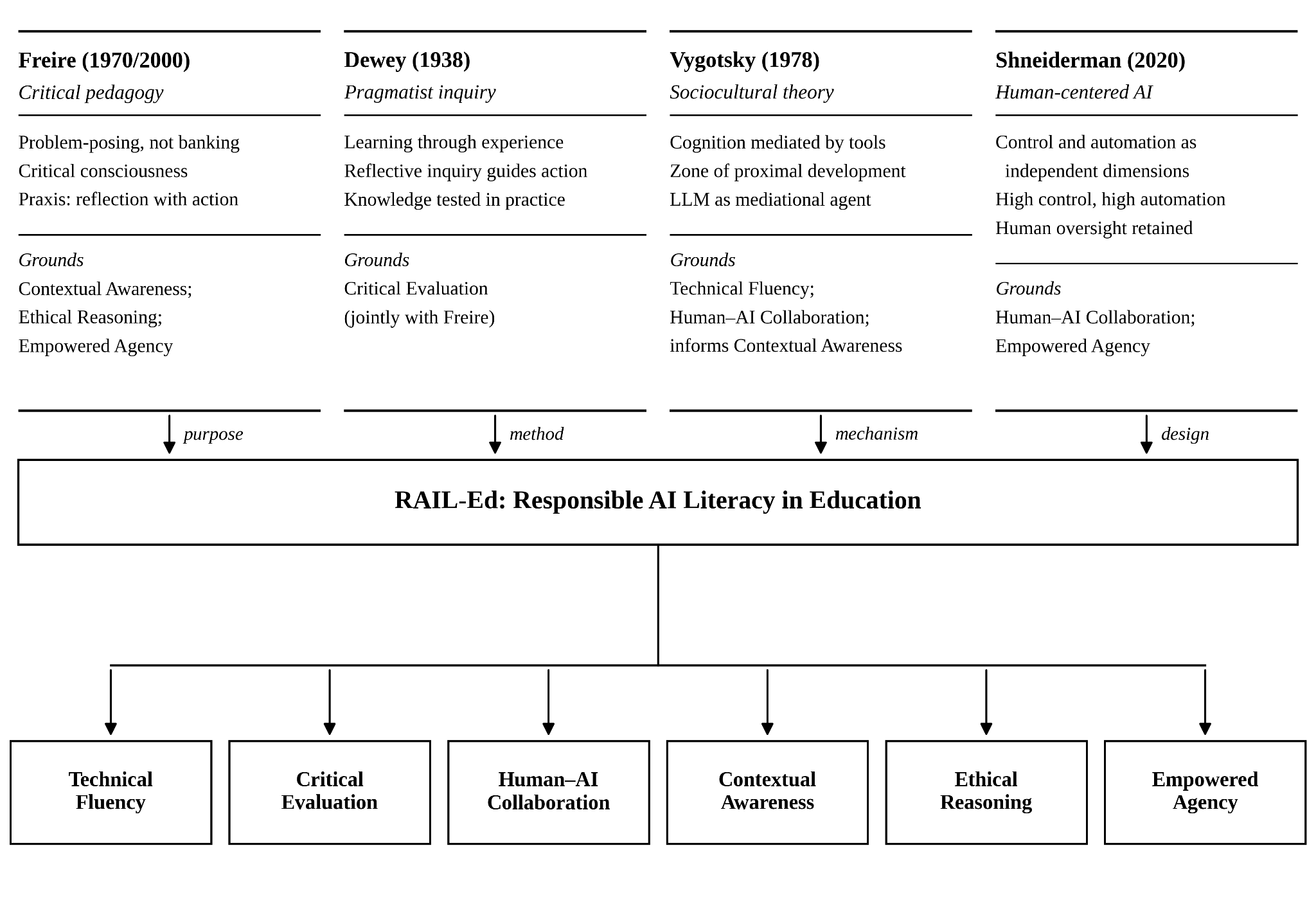}
\caption{Theoretical Foundations of RAIL-Ed Framework}
\label{fig:foundations}
\end{figure*}

\subsection{Existing AI Literacy and GenAI Literacy
Frameworks}\label{existing-ai-literacy-and-genai-literacy-frameworks}

Against these foundations, existing AI and GenAI literacy frameworks can
be read for what they offer and where they stop short. Long and Magerko
(2020) broadened AI literacy beyond technical knowledge, defining it as
a set of competencies enabling individuals to critically evaluate AI,
communicate and collaborate with it, and use it as a tool across
settings; their work established the human-centered turn on which later
frameworks build. Allen and Kendeou's (2024) ED-AI Lit framework
advanced an interdisciplinary, six-dimensional model of knowledge,
evaluation, collaboration, contextualization, autonomy, and ethics,
strengthening attention to independent judgment and contextual
implications. At the policy level, UNESCO's guidance (Miao et al., 2021)
foregrounded inclusion, equity, transparency, and human-centered
development, insisting that AI must not exacerbate existing
inequalities.

These contributions are foundational, but they share limitations for the
generative, teacher-preparation context. Most predate the widespread
adoption of LLMs and therefore emphasize foundational AI concepts over
the distinctive demands of generative systems: prompt construction,
hallucination detection, output verification, and human--AI co-creation
(Park, 2025). They tend to position educators and learners as users
rather than co-creators of AI-mediated content (Yim \& Su, 2025), and
they leave the ethics of generative use, academic integrity,
misinformation, copyright, privacy, algorithmic bias, and data
governance comparatively underdeveloped (Gruenhagen et al., 2024). A
teacher using GenAI to design assessments, for instance, must judge
whether generated items reproduce copyrighted material, encode bias, or
assert plausible but false content that undermines validity---judgments
these frameworks do not equip them to make. Recent work increasingly
characterizes GenAI literacy as a multidimensional competency spanning
understanding, effective use, critical evaluation, ethical reasoning,
and responsible creation (Park, 2025), but these conceptualizations have
not yet been consolidated into a theoretically grounded, teacher-focused
framework. Section 3 specifies these gaps in detail, and Table 2
situates RAIL-Ed among the full set of frameworks reviewed.

\section{Identified Gaps and the Rationale for a New
Framework}\label{identified-gaps-and-the-rationale-for-a-new-framework}

\subsection{Coverage of the Reviewed
Literature}\label{coverage-of-the-reviewed-literature}

To establish what a new framework must provide, we examined how the
existing literature allocates its attention and where it falls short in
the generative K--12 teacher-preparation context. We reviewed 67
publications on AI and GenAI literacy in education (2023--2025);
research addressing GenAI literacy specifically remains concentrated in
higher education, with K--12 teacher preparation comparatively
underserved. Across these studies, Long and Magerko's (2020)
design-centered model and Ng et al.'s (2021) four-domain construct are
the most frequently cited frameworks, both developed before large
language models were publicly released at scale, and neither designed
for the epistemic and pedagogical demands that tools such as ChatGPT,
Claude, and Gemini place on K--12 teachers.

To characterize where the literature concentrates, we coded each study
against the five analytic dimensions that structured our review:
knowledge, skill, ethics, equity, and agency (Walker \& Avant, 2005).
Because these dimensions were defined a priori from concept-analysis
methodology, the distribution in Table 1 reports what the corpus
contains independently of the framework it motivates. Coverage focuses
on knowledge and skills, while equity and agency receive substantive
treatment in a minority of studies, the systematic thinness that RAIL-Ed
is designed to address.

\begin{table*}[!tp]
\centering\small
\caption{Coverage of Five Analytic Dimensions Across the 67 Reviewed Studies (2023--2025)}
\begin{tabular}{@{}>{\RaggedRight\arraybackslash}p{\dimexpr 0.1741\textwidth-2\tabcolsep\relax}>{\RaggedRight\arraybackslash}p{\dimexpr 0.3892\textwidth-2\tabcolsep\relax}>{\RaggedRight\arraybackslash}p{\dimexpr 0.2167\textwidth-2\tabcolsep\relax}>{\RaggedRight\arraybackslash}p{\dimexpr 0.2200\textwidth-2\tabcolsep\relax}@{}}
\toprule
\textbf{Analytic dimension} & \textbf{What it captures} &
\textbf{Studies addressing it, n (\%)} & \textbf{Typical depth of
treatment} \\
\midrule
Knowledge & How AI and GenAI systems work; conceptual understanding & 65
(97\%) & Foundational \\
Skill & Operating, prompting, and applying tools & 64 (96\%) &
Substantive \\
Ethics & Integrity, bias, privacy, misinformation & 66 (99\%) & Present
but fragmented \\
Equity & Access, language, structural inequality & 49 (73\%) & Thin \\
Agency & Co-agency, principled use, restraint, civic voice & 40 (60\%) &
Thin/ instrumental \\
\bottomrule
\end{tabular}
\par\vspace{2pt}
\begin{minipage}{0.97\textwidth}\footnotesize \emph{Note. Dimensions follow the concept-analysis matrix used in coding (Walker \& Avant, 2005) and were defined prior to and independently of RAIL-Ed's pillars. Counts reflect studies giving each dimension substantive treatment; coding criteria and inter-coder agreement are reported in the companion review (Zhou et al., in press).}\end{minipage}
\end{table*}
The gaps below reflect this distribution. Running through them is a
single problem: existing frameworks, and even the most recent
international standards, specify what teachers and learners should be
able to do, but they do not provide a generative-specific, theoretically
grounded, and developmental account of how K--12 teachers acquire that
capacity. They name a destination without a theory of how to reach it.

\begin{table*}[!tp]
\centering\small
\caption{Comparison of Widely Cited AI and GenAI Literacy Frameworks}
\begin{tabular}{@{}>{\RaggedRight\arraybackslash}p{\dimexpr 0.1513\textwidth-2\tabcolsep\relax}>{\RaggedRight\arraybackslash}p{\dimexpr 0.1767\textwidth-2\tabcolsep\relax}>{\RaggedRight\arraybackslash}p{\dimexpr 0.2125\textwidth-2\tabcolsep\relax}>{\RaggedRight\arraybackslash}p{\dimexpr 0.1317\textwidth-2\tabcolsep\relax}>{\RaggedRight\arraybackslash}p{\dimexpr 0.1362\textwidth-2\tabcolsep\relax}>{\RaggedRight\arraybackslash}p{\dimexpr 0.1916\textwidth-2\tabcolsep\relax}@{}}
\toprule
\textbf{Framework (year)} & \textbf{Focus/definition} &
\textbf{Dimensions} & \textbf{Context} & \textbf{Audience} &
\textbf{Limitation for GenAI} \\
\midrule
Long \& Magerko (2020) & Human-centered AI literacy competencies & 17
competencies (recognize AI, critical evaluation, communication \&
collaboration, using AI as a tool) & General public & General learners &
Pre-LLM; prompting, hallucination, and output verification not
addressed \\
Ng et al. (2021) & Conceptual review of AI literacy & Four domains: know
\& understand, use \& apply, evaluate \& create, ethics & General
education & Learners & Predictive-AI framing; positions learners as
users, not co-creators \\
AI4K12 / Touretzky et al. (2019) & K--12 AI content standards & Five Big
Ideas (perception, representation \& reasoning, learning, natural
interaction, societal impact) & K--12 curriculum & K--12 students &
Content standards, not teacher GenAI literacy; no generative focus \\
UNESCO / Miao et al. (2021) & Global AI-in-education policy guidance &
Inclusion, equity, transparency, human-centered AI & International
policy & Policymakers & Policy-level, not a classroom competency
model \\
ED-AI Lit (Allen \& Kendeou, 2024) & Interdisciplinary AI literacy in
education & Knowledge, evaluation, collaboration, contextualization,
autonomy, ethics & Education (broad) & Educators \& learners & Not
generative-specific; equity not constitutive; not K--12
teacher-focused \\
Chiu et al. (2024) & AI literacy with ethics \& collaboration central &
Technology, Impact, Ethics, Collaboration, Self-Reflection & School
education & Students & Limited GenAI-specific competencies; no
developmental account \\
Su \& Yang (2023) --- IDEE & GenAI literacy / human-in-the-loop design &
Identify outcomes; Determine automation level; Ensure ethics; Evaluate
effectiveness & General teaching & Educators & Does not specify how
teachers develop the judgment it requires \\
Annapureddy et al. (2025) & GenAI competency model & 12 competencies
differentiating generative from predictive AI & Workforce/\allowbreak general &
General users & Generic; not K--12; no pedagogy or equity \\
Sattelmaier \& Pawlowski (2025) & AI competencies for K--12 teachers &
Teacher-facing AI competency areas & K--12 schools & K--12 teachers &
Limited equity / civic-agency treatment; emerging validation \\
UNESCO AI CFT (Miao \& Cukurova, 2024) & Global teacher AI competency
standard & 5 aspects (human-centred mindset; ethics of AI; AI
foundations \& applications; AI pedagogy; AI for professional
development) × 3 levels (Acquire, Deepen, Create) & K--12 teachers
(policy) & K--12 teachers & Competency standard; not a
generative-specific or developmental pedagogical model \\
UNESCO AI CF for Students (Miao et al., 2024) & Global student AI
competency standard & 4 aspects × 3 levels (Understand, Apply, Create);
12 competencies & K--12 curriculum & K--12 students & Student-facing;
not teacher preparation \\
OECD/EU AILit (2026) & AI literacy for primary \& secondary education &
4 domains: Engage with AI, Create with AI, Manage AI, Shape AI
(knowledge, skills, attitudes) & K--12 schooling & Learners
(teacher-mediated) & Learner-outcome framework; limited teacher
developmental progression \\
AI Literacy (Mills et al., 2024) & Practitioner framework (Digital
Promise) & Understand, evaluate, and use emerging AI & K--12 education &
Educators \& learners & Brief, practitioner-oriented; not
generative-pedagogy-specific \\
RAIL-Ed (this paper) & Responsible GenAI literacy for K--12 teacher
education & Six pillars: Technical Fluency, Critical Evaluation,
Human--AI Collaboration, Contextual Awareness, Ethical Reasoning,
Empowered Agency & K--12 teacher prep (pre- \& in-service) & K--12
teachers & Conceptual; awaiting empirical validation (stated openly) \\
\bottomrule
\end{tabular}
\par\vspace{2pt}
\begin{minipage}{0.97\textwidth}\footnotesize \emph{Note.} The table characterizes each framework's focus, dimensions, and principal limitation for GenAI; dimensions reflect each source's own terminology. The final row presents the framework proposed in this paper.\end{minipage}
\end{table*}
To make explicit how RAIL-Ed complements these standards, Table 3 maps
its six pillars onto the UNESCO AI Competency Framework for Teachers
(Miao \& Cukurova, 2024) and the OECD/EU AI literacy framework
(OECD/European Commission, 2026). RAIL-Ed operationalizes their shared
commitments for the generative, K--12 teacher-preparation context,
adding the developmental progression and dialectical account those
frameworks leave underspecified.

\subsection{Generative-Specificity}\label{generative-specificity}

As Annapureddy et al. (2025) indicated in their GenAI competency model,
which includes twelve items, existing AI literacy frameworks remain too
generic and fail to differentiate between predictive and GenAI systems.
This distinction shapes what GenAI literacy means and what it requires
teachers to be able to do. In another study, Sattelmaier and Pawlowski
(2025) observe that K--12 teachers are navigating rapid AI-driven
changes in their classrooms without a clear account of which
competencies they actually need to develop.

Building on the previous Gen AI literacy models, recent scholarship has
proposed frameworks that aim to integrate ethics, collaboration, and
interdisciplinary perspectives more centrally into AI literacy. For
example, Chiu et al. (2024) developed a five-component AI literacy
framework (Technology, Impact, Ethics, Collaboration, Self-Reflection)
that places ethics and collaboration at the center of the GenAI
framework. Similarly, Allen and Kendeou (2024) developed an ED-AI Lit
framework as a holistic, interdisciplinary model of AI literacy in
education. While these frameworks represent meaningful advances in
integrating interdisciplinary and ethical dimensions of AI literacy in
different subject domains, their contributions in GenAI literacy remain
constrained. Neither model addresses the generative-specific
competencies that distinguish GenAI literacy from broader AI literacy,
neither is designed for K--12 teachers, and neither treats equity as a
constitutive design principle. Cheah et al. (2025) further discussed
that the lack of clear policy and instructional guidance for integrating
GenAI in K--12 makes even motivated teachers ineffective in meeting
practical classroom needs in real time.

\subsection{Agency Narrowed to Tool
Adoption}\label{agency-narrowed-to-tool-adoption}

An important gap in the AI and GenAI literacy theoretical frameworks of
reviewed studies is the treatment of teacher and learner agency in the
case of GenAI systems. The concepts of agency across the reviewed
studies were consistently interpreted in narrow, instrumental terms
(Zhou et al., in press). Agency was defined as the behavioral intention of
an educator to adopt AI tools (e.g., Bower et al., 2024; Kong et al.,
2024), write effective prompts (Yang \& Appleget, 2025), and make
individual decisions about tool use (Cheah et al., 2025). This
definition is practical. However, it does not fully capture the ethical
and civic demands that scholars have argued must accompany GenAI
integration in education (Annapureddy et al., 2025; Laine et al., 2025;
Roe et al., 2025; Tagare et al., 2025). The current AI literacy
literature, and GenAI literacy studies specifically, has not yet
developed a theoretically grounded account of human--AI co-agency.
Although the OECD (2019) Learning Compass uses co-agency to describe
collaborative relationships among learners, teachers, peers, families,
and communities in support of student development, the concept has not
been systematically extended to human--AI relationships in the
educational literature. Studies that engage with human--AI collaboration
tend to use the related vocabulary of co-creation (MacDowell et al.,
2024; Yang \& Appleget, 2025) or distributed pedagogical roles (Yu \&
Pian, 2025), but do not articulate co-agency as a construct grounded in
a theory of learner and teacher participation in AI-mediated knowledge
production.

One negative consequence of this gap is blind reliance on GenAI tools.
Zhang et al. (2025) reported that pre-service teachers had an
over-reliance and less critical agency when they were practicing with
GenAI tools. Joseph (2023) provides a concrete illustration of why this
pattern is consequential. In her test of ChatGPT on a literary analysis
task, the model fabricated factual references that students would
otherwise have accepted as accurate (Joseph, 2023). Building on these
empirical findings, Jorolan et al. (2025) argued that AI dependency is a
crucial moral dimension that remains under-researched. Beyond these
empirical observations, existing frameworks have largely overlooked the
problem of blind reliance, treating it neither as a competency gap to
address nor as an ethical risk to design against.

\subsection{Ethics and Equity Treated as Compliance, Not
Design}\label{ethics-and-equity-treated-as-compliance-not-design}

At the framework level, Su and Yang (2023) propose that educators should
determine the appropriate level of automation when integrating GenAI
into teaching, a principle that aligns conceptually with the broader
human-in-the-loop (HITL) tradition in AI-in-education research (Memarian
\& Doleck, 2024), in which human judgment retains oversight over
AI-mediated processes and outputs. However, their GenAI literacy IDEE
framework (Identify the desired outcomes, Determine the appropriate
level of automation, Ensure ethical considerations, Evaluate the
effectiveness, Su \& Yang, 2023) does not specify the conditions under
which teachers develop the metacognitive capacity to exercise this
judgment consistently and critically. MacDowell et al. (2024) position
teachers as active creators who must develop the knowledge, skills, and
mindsets necessary to teach with and create using GenAI in pedagogically
responsible ways. This conception moves beyond tool adoption toward
sustained engagement and co-creation. However, it remains an
illustrative case, and it does not yet specify the conditions under
which such co-creative engagement develops at scale across teacher
preparation programs.

We argue that the field needs a theorization of co-agency that draws on
Shneiderman\textquotesingle s (2020) human-centered AI principles and
Freirean critical praxis (Freire, 1970/2000) alike. Such a theorization
would position teachers as neither passive consumers of AI outputs nor
uncritical opponents of AI tools, but instead as deliberate and
ethically-engaged co-participants in AI-mediated knowledge construction.
We see this dual grounding as necessary because human-centered design
alone does not address the critical consciousness that teachers and
students need to engage with AI on equitable terms.

The other problem is the risk of misuse of GenAI tools. This is
essential to how teachers and students use GenAI in classrooms. The most
frequently identified misuse risk across the corpus is academic
dishonesty (Bae et al., 2024; Bower et al., 2024; Bukar et al., 2024),
yet this concern, legitimate as it is, functions in most studies as a
compliance problem rather than as a literacy challenge. The deeper
misuse risks of GenAI, such as hallucination and confabulation (Bae et
al., 2024; Choi, 2025), representational bias embedded in training data
(Feldman-Maggor et al., 2025), and the epistemic risk of AI-generated
content that mimics authoritative knowledge while being factually
unreliable, receive far less systematic attention and almost no
pedagogical operationalization in the included frameworks.
Feldman-Maggor et al. (2025) identified hallucination and gender-racial
representational bias in ChatGPT outputs as challenges for chemistry
teacher education, showing that disciplinary content expertise is
necessary for teachers to detect content errors and fabricated
references, yet insufficient for recognizing representational bias,
which demands AI-specific competencies beyond existing teacher-knowledge
frameworks. This finding has broad implications for K--12 teacher
preparation across subjects.

\subsection{Prompt Literacy Left
Unnamed}\label{prompt-literacy-left-unnamed}

Prompt literacy, which is an essential competency to construct
purposeful, critically reflective, and pedagogically intentional inputs
for LLMs, represents another concerning issue. Wang et al. (2025)
identified prompt literacy as an emergent but underdeveloped competency
that existing frameworks do not meaningfully address. A reasonable
response might be that prompt literacy could be located within the
knowledge or skills components of existing AI literacy frameworks: Ng et
al.\textquotesingle s (2021) "use and apply" domain, Long and
Magerko\textquotesingle s (2020) design-centered competencies, or Allen
and Kendeou\textquotesingle s (2024) Knowledge component could in
principle accommodate it. In practice, however, none of these frameworks
explicitly names prompt construction as a competency, operationalizes
how teachers should develop it, or distinguishes it from general AI tool
use. A general "knowledge of AI" or "use of AI" component is different
from a framework that specifies prompt literacy as a constitutive
competency of GenAI literacy.

Prompt literacy is the operational bridge between teacher intention and
AI output, and the field has documented that teachers without explicit
prompt literacy instruction tend to default to broad, unstructured
prompts that constrain the quality of what AI returns. Yang and Appleget
(2025), for instance, identified three levels of pre-service-teacher
prompting and found that most prompts were broad and not purposefully
structured. Arrington et al. (2025) proposed a four-component prompt
engineering framework for K--12 teachers. Barbieri and Nguyen (2025)
found that pre-service teachers who received structured prompt training
demonstrated significantly higher self-efficacy and more purposeful AI
use during practicum placements. Yet Şimşek (2025) documented
considerable variation in the quality and intentionality of
teacher-generated prompts among practicing and pre-service mathematics
teachers. This is the persistent distance between understanding GenAI in
theory and orchestrating it purposefully in classroom practice. This gap
cannot be closed by frameworks that simply enumerate competencies. What
is missing is a clear account of what purposeful prompt construction
looks like across different pedagogical contexts. Hence, we believe that
a new framework must treat prompt literacy not as an ancillary skill but
as a foundational dimension of GenAI literacy.

\subsection{The Alignment Question: Why a New Framework Is Still
Needed}\label{the-alignment-question-why-a-new-framework-is-still-needed}

Two international frameworks released since 2024 are especially
pertinent. UNESCO's AI Competency Framework for Teachers (Miao \&
Cukurova, 2024) specifies fifteen competencies across five aspects,
progressing through three levels (Acquire, Deepen, Create); the
OECD/European Commission's Empowering Learners for the Age of AI (2026)
organizes AI literacy as a developmental pathway across four domains:
Engage, Create, Manage, and Shape AI. Together they establish an
authoritative, human-centered consensus that AI literacy is constituted
by ethical and civic capacities, not technical skill alone (cf. Mills et
al., 2024). RAIL-Ed's alignment with them is deliberate, and it is not
redundant. These standards specify what teachers and learners should be
able to do; they neither theorize the generative-specific dynamics that
determine whether a given affordance deepens or displaces learning, nor
supply a developmental account of how teachers acquire that capacity.
UNESCO's framework is a competency standard, and the OECD/EU framework
is a learner-outcome pathway with limited teacher developmental
progression.

To make explicit how RAIL-Ed complements these standards, Table 3 maps
its six pillars onto the UNESCO AI Competency Framework for Teachers
(Miao \& Cukurova, 2024) and the OECD/EU AI literacy framework
(OECD/European Commission, 2026). RAIL-Ed operationalizes their shared
commitments for the generative, K--12 teacher-preparation context,
adding the developmental progression, the dialectical account of
affordances, and the constitutive treatment of equity that those
frameworks leave underspecified.

\begin{table*}[!tp]
\centering\small
\caption{Alignment of RAIL-Ed Pillars With the UNESCO and OECD/EU Frameworks}
\begin{tabular}{@{}>{\RaggedRight\arraybackslash}p{\dimexpr 0.2328\textwidth-2\tabcolsep\relax}>{\RaggedRight\arraybackslash}p{\dimexpr 0.2312\textwidth-2\tabcolsep\relax}>{\RaggedRight\arraybackslash}p{\dimexpr 0.2027\textwidth-2\tabcolsep\relax}>{\RaggedRight\arraybackslash}p{\dimexpr 0.3333\textwidth-2\tabcolsep\relax}@{}}
\toprule
\textbf{RAIL-Ed pillars} & \textbf{UNESCO AI CFT aspect} &
\textbf{OECD/EU AILit domain} & \textbf{What RAIL-Ed adds for K--12
teachers} \\
\midrule
Technical Fluency; Critical Evaluation & AI foundations \& applications
& Engage with AI & Generative-specific mechanics; prompting as
research-design rehearsal; Emerging--Competent--Advanced progression \\
Human--AI Collaboration; Contextual Awareness & AI pedagogy & Create
with AI; Manage AI & Co-agency as a construct; situated,
equity-conscious adaptation as a classroom competency \\
Ethical Reasoning; Empowered Agency & Human-centered mindset; Ethics of
AI & Shape AI & Tier-differentiated integrity; principled non-use;
dialectical account of risk and benefit \\
\bottomrule
\end{tabular}
\par\vspace{2pt}
\begin{minipage}{0.97\textwidth}\footnotesize \emph{Note.} UNESCO aspects from Miao and Cukurova (2024); OECD/EU domains from OECD/European Commission (2026). The CFT's fifth aspect, AI for professional development, is a delivery mechanism spanning all six pillars rather than mapping to any single pillar. RAIL-Ed operationalizes these standards for K--12 teacher preparation; it does not replace them.\end{minipage}
\end{table*}
\subsection{Rationale for RAIL-Ed
Framework}\label{rationale-for-rail-ed-framework}

Our proposed framework, RAIL-Ed, responds to these converging demands.
Its six interdependent pillars are not a sequence of competencies but
mutually constituting dimensions of a single literacy orientation,
synthesizing Freirean critical pedagogy, Deweyan inquiry, Vygotskian
sociocultural theory, and Shneiderman's human-centered design (Dewey,
1938; Freire, 1970/2000; Shneiderman, 2020; Vygotsky, 1978). It is
responsive to international standards, the UNESCO AI Competency
Framework for Teachers (Miao \& Cukurova, 2024) and the OECD/EU AI
literacy framework (OECD/European Commission, 2026), and to the
governance commitments of the OECD AI Principles (OECD, 2024) and the EU
AI Act (European Union, 2024). We do not claim to resolve the field's
fragmentation by assertion; we offer a theoretically integrated
architecture from which cumulative development of the framework can
proceed.

\section{The RAIL-Ed Framework}\label{the-rail-ed-framework}

\subsection{Framework Overview}\label{framework-overview}

The RAIL-Ed framework synthesizes the gaps documented in Section 3 into
an integrative, justice-centered architecture for GenAI literacy across
K--12 teacher education, spanning pre-service and in-service teachers.
It is built around six interdependent pillars (Technical Fluency,
Critical Evaluation, Human--AI Collaboration, Contextual Awareness,
Ethical Reasoning, and Empowered Agency).

We define GenAI literacy for teachers as the integrated capacity of
teachers to understand how generative systems produce outputs, to
evaluate those outputs critically, to collaborate with them in
pedagogically and ethically responsible ways, and to exercise principled
agency, including restraint, over their role in learning. This
definition deliberately extends Long and Magerko's (2020) canonical
account of AI literacy as ``a set of competencies that enables
individuals to critically evaluate AI technologies'' (p. 2) into the
generative context, where evaluation must contend with fluent
fabrication and collaboration must preserve pedagogical authority. The
six pillars operationalize this definition; each is defined below and
anchored to the theoretical tradition from which it draws.

\emph{Technical Fluency} is the capacity to understand how generative
systems work, how transformer-based models are trained, how
reinforcement learning from human feedback shapes their behavior, and
why hallucination, bias propagation, and stochasticity are structural,
and to translate that understanding into purposeful prompt construction.
Grounded in Vygotsky's (1978) account of mediated cognition, the pillar
treats the generative model as a cultural tool that mediates thinking: a
teacher who does not grasp how the tool shapes its outputs is mediated
by it. It requires teachers to distinguish generative from predictive
systems (Annapureddy et al., 2025), to articulate in age-appropriate
terms why fluent prose is not evidence of truth, and to approach
prompting as research-design rehearsal.

\emph{Critical Evaluation} is the ability to critically assess AI
outputs for accuracy, bias, and source credibility. Grounded in Freire's
(1970/2000) critical consciousness and Dewey's (1938) reflective
inquiry, it reframes evaluation from a one-time accuracy check into a
habitual, dialogic stance toward AI-mediated knowledge, asking not only
whether an output is correct but whose knowledge it centers and whose it
omits. It requires teachers to triangulate AI claims against
disciplinary sources, to detect fabricated citations and
representational bias (Feldman-Maggor et al., 2025), and to model that
scrutiny so students internalize it as practice.

\emph{Human--AI Collaboration} is the capacity to engage generative
systems as contested co-participants in teaching and learning,
negotiating, co-authoring, and iterating with them while preserving
pedagogical authority. Grounded in Shneiderman's (2020) human-centered
AI, which holds that high human control and high automation are
independent dimensions to be maximized together, the pillar positions
the teacher as retaining oversight precisely while leveraging the
system's generativity. Vygotsky's (1978) mediation extends this to the
classroom, where the model functions as a mediational agent in lesson
design and inquiry, participating in the work without occupying the
social position of a partner (Tate et al., 2026). It requires teachers
to plan with AI to sequence scaffold-removal across assignments, and to
name and resist voice homogenization (Sourati et al., 2026; Warschauer
et al., 2023).

\emph{Contextual Awareness} is the capacity to situate GenAI within the
institutional, cultural, linguistic, and political conditions in which
teaching occurs, recognizing that the same tool produces different
consequences across differently resourced settings. Grounded in Freire's
(1970/2000) attention to the sociopolitical conditions of knowledge and
in sociocultural theory's insistence that cognition is situated, the
pillar treats questions of access, power, and whose knowledge is
centered as constitutive. It requires teachers to critically examine AI
models and data sources, to adapt activities to learners' contexts, and
to anticipate differential affordance outcomes (Cheah et al., 2025).

\emph{Ethical Reasoning} is the capacity to address bias, privacy,
authorship, and academic integrity as connected ethical commitments.
Drawing on Freire (1970/2000), this pillar conceptualizes ethics as
integral to education. It distinguishes compliance-based,
norm-dependent, and principled moral reasoning, emphasizing teachers'
role in fostering higher levels of moral development. It requires
teachers to recognize which tier a learner reasons from, to design
conditions that support principled integrity, and to treat transparency
about their own AI use: prompt logs, model versions, provenance as
responsible epistemic stewardship.

\emph{Empowered Agency} is the capacity to exercise principled,
self-authored judgment over GenAI, deciding when to adopt, when to
restrain, and when to refuse, and to extend that judgment into civic
voice and the transformation of AI systems. Grounded in Freire's
(1970/2000) praxis, in which reflection and action combine to transform
conditions, and in Shneiderman's (2020) insistence on sustained human
control, the pillar treats principled non-use as a sophisticated
literacy posture. It requires teachers to sustain reflective practice
under situational pressure, to participate in AI governance rather than
accept it passively, and to cultivate the same agency in their students,
with participation in design justice as its culminating expression.

RAIL-Ed is distinguished from existing models by three architectural
commitments. It is integrative, recognizing that the absence of any
single pillar creates specific pedagogical failure modes (Table 5). This
is the framework's central testable proposition: if development in one
pillar were shown to compensate fully for the absence of another, if,
for example, advanced technical fluency alone eliminated the failure
modes associated with absent critical evaluation, the integrative
commitment would be falsified. It is developmental, with each pillar
advancing through Emerging, Competent, and Advanced levels (Table 6). It
is also dialectical, acknowledging that generative AI affordances
simultaneously enable and constrain learning: the same capabilities that
support source discovery can produce fabricated information (the source
discovery--fabrication paradox), and the efficiency gained through AI
use can undermine the development of underlying skills. Thus, the
outcomes of AI engagement depend on the literacies users bring to these
interactions (cf. affordance theory; Gibson, 1979).

This dialectical commitment is what allows RAIL-Ed to address the
discipline of education's dual mandate with respect to GenAI: not to
choose between guarding against its risks and harnessing its potential,
but to cultivate the literacies that determine which of the two a shared
affordance produces. Generative systems can deepen inquiry, widen access
for multilingual and underprepared learners, and make expert reasoning
visible; the same systems can displace cognition, homogenize voice, and
naturalize bias. RAIL-Ed treats these not as a ledger to be balanced but
as conditional outcomes of educator and learner capacity, and it
specifies the capacities accordingly. In doing so, the framework
positions GenAI as a contested epistemic co-participant in the work of
teaching and learning rather than as either a threat to be contained or
a shortcut to be exploited.

Methodologically, the six pillars were derived through a hybrid
systematic review (a PRISMA-style search and screening) and qualitative
framework analysis, reported in full in a companion review (Zhou et al., in press) and summarized here. From an initial pool of 934 records (ERIC,
Web of Science, ScienceDirect, Scopus, IEEE Xplore, and ProQuest;
2023--2025), 67 studies met the inclusion criteria and were coded
against the five analytic dimensions of the concept-analysis matrix:
knowledge, skill, ethics, equity, and agency (Walker \& Avant, 2005).
The six pillars are not a relabeling of these five dimensions but a
theoretically motivated restructuring of them into teacher capacities,
governed by three criteria: each pillar must name a capacity whose
absence produces a distinct, documented pedagogical failure (Table 5);
each must admit a distinct developmental trajectory (Table 6); and each
must be groundable in the theoretical traditions established in Section
2. Applying these criteria produced two departures from the coding
matrix.

First, the skill dimension resolved into two pillars, Critical
Evaluation and Human--AI Collaboration, because the generative context
splits what the corpus treats as one competence: judging outputs that
may be fluently fabricated and working with a system as a co-participant
are distinct capacities with distinct failure modes (unverified error
versus displaced cognition). Second, equity was recast from a coverage
category into a constitutive commitment: rather than standing as a
separate pillar that could be satisfied by topical mention, it is built
into Contextual Awareness as the capacity to anticipate differential
consequences across differently resourced settings, and it recurs as a
design criterion across the remaining pillars. Knowledge maps onto
Technical Fluency, ethics onto Ethical Reasoning, and agency onto
Empowered Agency; in each case, the literature\textquotesingle s content
is reframed into the judgment the teacher exercises.

\subsection{Visual Representation of the
Framework}\label{visual-representation-of-the-framework}

Figure 2 renders RAIL-Ed as six interdependent pillars arranged around a
central hub representing GenAI literacy for teachers, the integrated
capacity defined in Section 4.1. The hub is encircled by a ring carrying
the framework\textquotesingle s three architectural commitments,
integrative, developmental, and dialectical, which describe how the
pillars work together. Read functionally, and in a grouping distinct
from the dimensional mapping developed in Section 4.3, the pillars play
three roles: Technical Fluency, Critical Evaluation, and Human--AI
Collaboration supply the cognitive capacities a teacher draws on at the
point of decision; Contextual Awareness and Ethical Reasoning supply the
capacity to read the institutional, cultural, and moral conditions
within which those decisions are made; and Empowered Agency supplies the
disposition that sustains principled practice across changing
conditions. The pillars are mutually constituting, which the figure
signals by connecting each pillar to the hub and, along the dashed ring,
to its neighbors. To illustrate: a teacher who notices a fabricated
citation in an AI-generated reading list (Critical Evaluation) revises
the prompt that produced it (Technical Fluency), decides with the class
how the tool should be used (Human--AI Collaboration), weighs what the
error means for the particular students in the room (Contextual
Awareness and Ethical Reasoning), and models the judgment students are
meant to develop (Empowered Agency), a single decision moving through
all six pillars at once. Table 5 formalizes the converse: the
characteristic failures that follow when any one of these capacities is
absent.

\begin{figure}[t]
\centering
\includegraphics[width=\columnwidth]{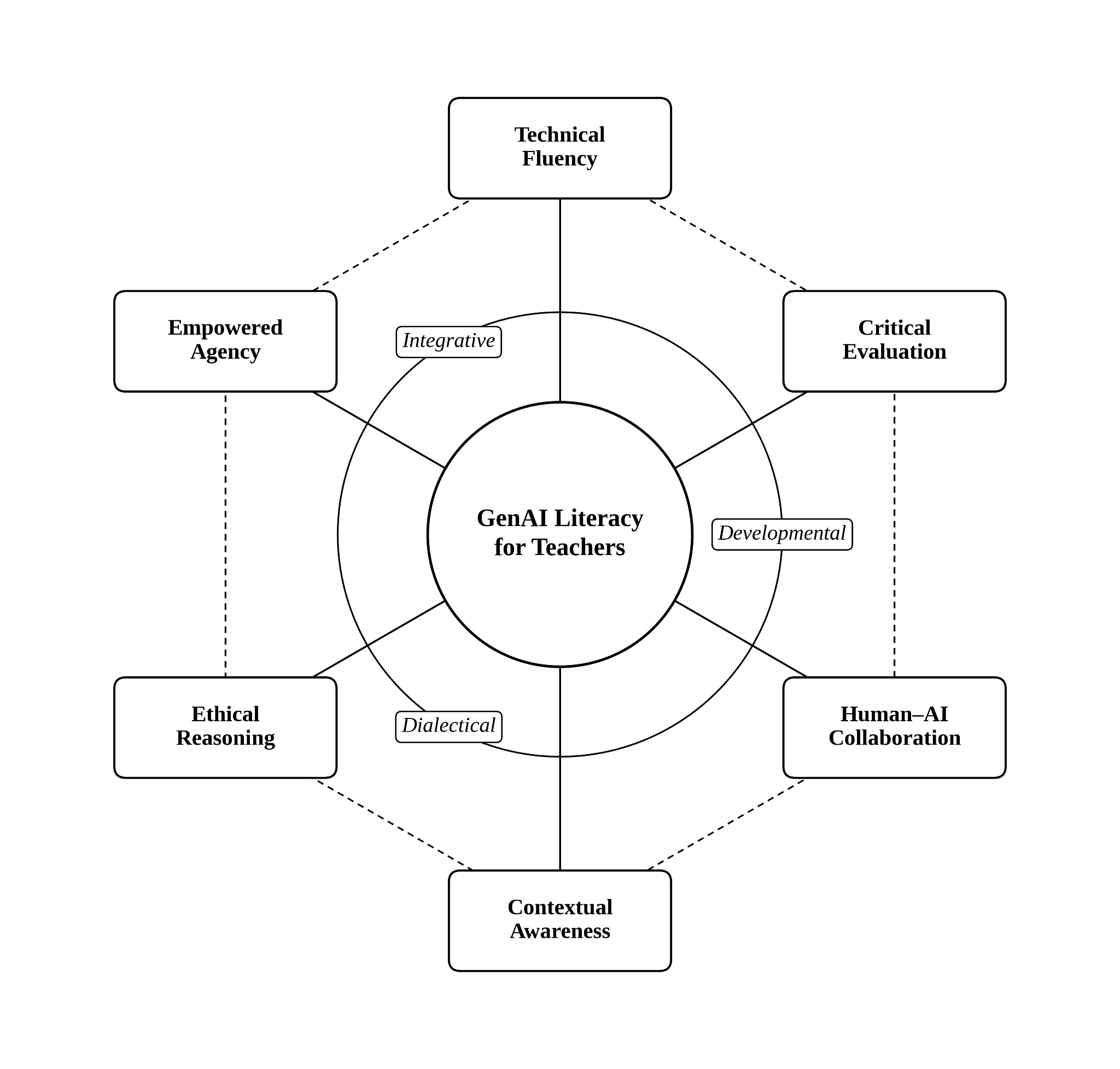}
\caption{The Six Pillars of RAIL-Ed: An Integrative Model of GenAI Literacy
for Teachers}
\label{fig:pillars}
\end{figure}

\subsection{Key Components and
Dimensions}\label{key-components-and-dimensions}

RAIL-Ed organizes GenAI literacy along four interlocking dimensions:
Knowledge, Skill, Ethics, and Critical Dispositions, operationalized
through the six pillars. Where Section 4.2 grouped the pillars by their
function at the point of decision, this section maps them by the kind of
capacity each develops; Table 4 provides the crosswalk between each
pillar, the gap it addresses, the core competencies and GenAI-specific
content it requires, and the pedagogical praxes through which it is
enacted.

\subsubsection{Knowledge Dimensions}\label{knowledge-dimensions}

\textbf{AI Basics.} Technical Fluency begins from a premise about
teaching, not about content: teachers cannot teach or assess what they
do not themselves understand (Shulman, 1986). AI4K12's Five Big Ideas
(Touretzky et al., 2019)- perception, representation and reasoning,
learning, natural interaction, and societal impact- specify what K--12
students should understand about AI; RAIL-Ed does not adopt that scope
as its own but treats teacher mastery of it as the floor beneath
Technical Fluency. The link is pedagogical, not curricular: AI4K12
defines the student-facing content, while RAIL-Ed defines the teacher
capacity required to make that content teachable and to evaluate
students' grasp of it. Two things distinguish the teacher's capacity
from the student's content. First, RAIL-Ed extends these largely pre-LLM
ideas to the distinction AI4K12 never draws between predictive and
generative systems (Annapureddy et al., 2025), because the phenomena
that matter most for GenAI literacy (hallucination, stochasticity,
fluent-but-false output) are properties of generation, not of AI in
general. Second, it reframes the Big Ideas from content to be covered
into judgment to be exercised: a teacher with Technical Fluency uses
this understanding to explain, in age-appropriate terms, why an LLM does
not retrieve facts the way a search engine does and why fluent prose is
not evidence of truth, and models that reasoning for students.

\textbf{GenAI Mechanics.} Educators and learners require functional
knowledge of how transformer-based LLMs are trained, how reinforcement
learning from human feedback (RLHF) shapes their behavior, and why these
mechanics make hallucination, bias propagation, and stochasticity
structural phenomena. This pillar responds to Feldman-Maggor et al.'s
(2025) demonstration that disciplinary expertise, while necessary for
detecting content errors and fabricated references, is not sufficient:
in their gender-bias example, ``the teachers' TPACK was not enough to
critically evaluate the chat output'' (p. 7), because recognizing
representational bias requires AI-specific knowledge that existing
teacher-knowledge frameworks do not capture. RAIL-Ed therefore treats
mechanics literacy as a prerequisite for evaluation literacy: a teacher
who understands why a model generates fluent falsehoods is positioned to
evaluate outputs that disciplinary knowledge alone cannot adjudicate.

\textbf{Prompt Engineering.} Where existing frameworks treat prompting
as ancillary, RAIL-Ed positions it as foundational, consistent with the
intervention evidence in Arrington et al. (2025) and with emerging
measurement work: Yan et al. (2026) found prompt-crafting so central to
competent use that they added it to the usage dimension of their
validated GenAI literacy scale, identifying it as a component
``previously overlooked in existing scales'' (p. 12). Prompt engineering
is conceived as research-design rehearsal: the construction of an
epistemically front-loaded query in which theoretical commitments,
audience, and disciplinary conventions are made explicit before the
model is engaged. Prompt literacy is neither intuitive nor automatically
acquired through tool exposure (Şimşek, 2025; Wang et al., 2025) and
must be taught explicitly.

\subsubsection{The Skill Dimension}\label{the-skill-dimension}

\textbf{Educational Planning.} Educators must design lessons in which
GenAI is integrated as a contested co-participant. This involves
selecting tasks where AI augments, not replaces, cognitive labor and
sequencing assignments along a scaffold-removal logic that proceeds from
AI-permitted brainstorming to independent drafting to closed-environment
demonstration. The pillar addresses Yu and Pian's (2025) documented
knowing--doing gap: the persistent distance between understanding GenAI
in theory and orchestrating it purposefully in practice.

\textbf{Assessment Design.} Assessment in an LLM-saturated environment
requires process-visible architectures that surface how a learner
arrived at an outcome, not only what the outcome looks like. Drawing on
Bower et al.'s (2024) finding that educators prioritize process-focused
assessment in response to generative AI, this competency includes
revision logs and reflection journals; tiered tasks distinguishing
incidental, substantive, and generative use (Hossain, 2026a); and prompts
that LLMs cannot plausibly complete without disciplinary triangulation.
Consistent with the developmental account of integrity developed below,
detection-based assessment is treated as architecturally incapable of
producing durable academic integrity (Hossain, 2026a).

\textbf{Communication and Collaboration.} RAIL-Ed treats communication
as both interpersonal and epistemic: communication with students and
colleagues about responsible use, and communication with the AI system
itself as a documented, citable participant in knowledge production.
Voice homogenization (Sourati et al., 2026) is treated as a
collaboration risk to be named and resisted, especially for multilingual
learners (Warschauer et al., 2023).

\subsubsection{Ethical Dimensions}\label{ethical-dimensions}

\textbf{Bias Awareness.} Educators must recognize that training-data
composition encodes representational asymmetries (gender, racial,
linguistic, disciplinary) that surface in classroom-relevant ways:
stereotyped examples, narrowed canons, and outputs that perform fluency
while obscuring whose knowledge has been centered. Every AI output is
interrogated for what is present, what is absent, and whose authority
the absence reflects (Feldman-Maggor et al., 2025).

\textbf{Privacy and Transparency.} Privacy literacy in RAIL-Ed is
operational. Educators must know what data the platforms they use
collect, what their districts and institutions permit, and what they are
obligated to disclose (Miao \& Holmes, 2023). Transparency literacy
extends to one's own practice: RAIL-Ed treats prompt logs, model
versions, and dataset provenance as responsible epistemic stewardship.

\textbf{Academic Integrity Considerations.} Academic integrity in
RAIL-Ed is treated developmentally. Drawing on Kohlberg's (1984) stages
of moral development, the framework's three-tier model (Hossain, 2026a)
distinguishes three structurally distinct tiers of ethical reasoning,
compliance-based (governed by fear of detection), norm-dependent
(governed by instructor authority), and principled (governed by
self-authored values), that produce different failure modes, such that
one-size-fits-all detection policy is architecturally incapable of
cultivating principled reasoning. The competency for educators is to
recognize which tier a learner is reasoning from and to design
conditions that support upward developmental movement.

\subsubsection{Critical
Dispositions}\label{critical-dispositions}

\textbf{Reflective Thinking.} Reflective thinking extends Dewey's (1938)
inquiry-based stance into the AI-mediated environment. Educators and
learners are asked, recurrently, what intellectual work AI performed,
displaced, or privileged in a given encounter and what that means for
the development of their own judgment. Reflection is not an end-of-term
exercise but a structural feature of practice.

\textbf{Ethical Decision-Making.} Ethical decision-making is the
disposition through which knowledge, skill, and ethical reasoning are
translated into action under situational pressure, a deadline at
midnight, a high-stakes assignment, a politicized classroom topic,
pressures the prevailing literature has not theorized adequately.
Drawing on Freirean praxis (Freire, 1970/2000) and Shneiderman's (2020)
human-centered AI principles, this disposition extends agency beyond
individual restraint to include civic voice, participatory critique, and
creative transformation of AI systems. RAIL-Ed treats principled non-use
as a legitimate outcome of AI literacy rather than a literacy failure:
an informed, research-based refusal that constitutes a sophisticated
literacy posture (Hossain, 2026a). Design-justice participation is
treated as the framework's culminating expression of empowered agency.

\begin{table*}[!tp]
\centering\small
\caption{RAIL-Ed Pillars: Gaps Addressed, Competencies, GenAI Content, and Pedagogical Praxes}
\begin{tabular}{@{}>{\RaggedRight\arraybackslash}p{\dimexpr 0.2109\textwidth-2\tabcolsep\relax}>{\RaggedRight\arraybackslash}p{\dimexpr 0.2735\textwidth-2\tabcolsep\relax}>{\RaggedRight\arraybackslash}p{\dimexpr 0.2817\textwidth-2\tabcolsep\relax}>{\RaggedRight\arraybackslash}p{\dimexpr 0.2339\textwidth-2\tabcolsep\relax}@{}}
\toprule
\textbf{RAIL-Ed pillar (dimension)} & \textbf{Deficit addressed} &
\textbf{Core competencies and GenAI-specific content} &
\textbf{Pedagogical praxes} \\
\midrule
Technical Fluency (Knowledge) & Prompt literacy treated as ancillary;
predictive-vs-GenAI conflation (Annapureddy et al., 2025). & LLM
training and RLHF; hallucination and stochasticity; prompt engineering
as research-design rehearsal. & Model demonstrations; tool
deconstruction; prompt-revision logs. \\
Critical Evaluation (Skill) & Frameworks list ``evaluation'' as a
competency without specifying disciplinary application (Sperling et al.,
2024). & Bias audit; fabricated-citation detection; disciplinary
triangulation; deepfake analysis. & Fact-checking workshops; bias
audits; verification protocols. \\
Human--AI Collaboration (Skill) & Agency narrowed to tool-adoption
intention; human-in-the-loop under-operationalized (Su \& Yang, 2023). &
Ethical co-engagement; co-authorship; prompt iteration; GenAI-assisted
lesson design. & Case-based PD; design challenges; scaffold-removal
sequencing. \\
Contextual Awareness (Knowledge--Skill bridge) & Frameworks
decontextualized from K--12 vs. HE realities; weak attention to power
and language (Cheah \& Kim, 2025). & Political economy of foundation
models; data colonialism; AI governance; linguistic adaptation. &
Sociocultural inquiry; policy analysis; multilingual comparison. \\
Ethical Reasoning (Ethics) & Ethics treated as compliance checklist;
bias, privacy, integrity addressed in isolation (Tagare et al., 2026). &
Authorship and surveillance; LLM training-data privacy;
tier-differentiated integrity. & Ethical scenarios; phronesis-based
reflection; developmental disclosure. \\
Empowered Agency (Critical Dispositions) & Affective and civic
dimensions absent or reduced to technology acceptance. & Civic voice;
participatory critique; creative transformation; principled non-use. &
Participatory design; advocacy projects; design-justice activities. \\
\bottomrule
\end{tabular}
\par\vspace{2pt}
\begin{minipage}{0.97\textwidth}\footnotesize \emph{Note.} Gaps are documented in the systematic review reported in Section 3. Pedagogical praxes are illustrative. PD = professional development.\end{minipage}
\end{table*}
\subsubsection{The Failure-Mode Map: Why Each Pillar Is
Load-Bearing}\label{the-failure-mode-map-why-each-pillar-is-load-bearing}

RAIL-Ed's integrative commitment, the claim that the six pillars are
mutually constituting and jointly necessary, is operationalized in Table
5 as a failure-mode map. Each single-pillar row pairs a capacity with
the pedagogical failure documented in the reviewed literature when that
capacity is absent; the three compound rows state the framework's own
predictions for what follows when pairs of pillars are missing. The map
is the empirical face of the testable proposition stated in Section 4.1.
Each row predicts a deficit that development in the remaining pillars
should not, on the framework's account, remediate. The catastrophic
condition at the foot of the table names the field's current default,
AI-saturated learning without AI literacy (Cheah \& Kim, 2025), and
RAIL-Ed exists to make this default unacceptable.

\begin{table*}[!tp]
\centering\small
\caption{Failure-Mode Map: Documented Consequences of Absent or Underdeveloped Pillars}
\begin{tabular}{@{}>{\RaggedRight\arraybackslash}p{\dimexpr 0.3262\textwidth-2\tabcolsep\relax}>{\RaggedRight\arraybackslash}p{\dimexpr 0.6738\textwidth-2\tabcolsep\relax}@{}}
\toprule
\textbf{Absent or underdeveloped capacity} & \textbf{Resulting
pedagogical failure mode} \\
\midrule
Technical Fluency & Users conflate LLMs with search engines and read
fluent prose as evidence of truth, leaving hallucination unanticipated
(the LLM-as-oracle error). \\
Critical Evaluation & Fabricated citations, biased framings, and
unverified claims pass unchecked into student and teacher work. \\
Human--AI Collaboration & AI is either banned outright or used as an
unsupervised substitute; cognitive labor is displaced rather than
scaffolded. \\
Contextual Awareness & Activities imported from well-resourced settings
deepen access and language asymmetries; questions of whose knowledge is
centered go unasked. \\
Ethical Reasoning & Integrity collapses into detection; learners comply
under surveillance but internalize no principle (compliance theater). \\
Empowered Agency & Adoption proceeds by peer default or reflexive
refusal; no durable, self-authored stance survives changing
conditions. \\
Technical Fluency + Critical Evaluation (compound) & Fluent dependence:
confident, well-prompted use of output that is never verified. \\
Human--AI Collaboration + Critical Evaluation (compound) & Confident
error: smoothly integrated work built on fabricated or biased
material. \\
Critical Evaluation + Technical Fluency / prompt literacy (compound) &
Degraded scrutiny: careful checking of outputs already compromised by
the questions that produced them. \\
All pillars absent (catastrophic) & AI-saturated learning without AI
literacy---the field's current default condition (Cheah \& Kim,
2025). \\
\bottomrule
\end{tabular}
\par\vspace{2pt}
\begin{minipage}{0.97\textwidth}\footnotesize \emph{Note.} Single-pillar failure modes synthesize gaps documented in the Section 3 review; compound failure modes are the framework's predictions, stated as testable claims.\end{minipage}
\end{table*}
\subsubsection{The Developmental Rubric: Emerging, Competent,
Advanced}\label{the-developmental-rubric-emerging-competent-advanced}

Table 6 specifies Emerging, Competent, and Advanced levels for each
pillar against which educators, teacher-preparation programs, and
accrediting bodies can assess maturation. Two design principles
distinguish this rubric from existing developmental schemes. First,
advancement in one pillar does not compensate for stagnation in another:
an educator who is Advanced in Technical Fluency but Emerging in Ethical
Reasoning produces a compound failure, not a partial success. Second,
advancement is recursive: encounters with new AI generations, new
student populations, or new policy environments can require
recalibration in specific pillars, a feature the rubric treats as
healthy practice.

\begin{table*}[!tp]
\centering\small
\caption{Developmental Rubric for the Six Pillars of RAIL-Ed}
\begin{tabular}{@{}>{\RaggedRight\arraybackslash}p{\dimexpr 0.1730\textwidth-2\tabcolsep\relax}>{\RaggedRight\arraybackslash}p{\dimexpr 0.2669\textwidth-2\tabcolsep\relax}>{\RaggedRight\arraybackslash}p{\dimexpr 0.2718\textwidth-2\tabcolsep\relax}>{\RaggedRight\arraybackslash}p{\dimexpr 0.2883\textwidth-2\tabcolsep\relax}@{}}
\toprule
\textbf{Pillar} & \textbf{Emerging} & \textbf{Competent} &
\textbf{Advanced} \\
\midrule
Technical Fluency & Uses default prompts; conflates LLMs with search
engines; limited awareness of hallucination. & Designs purposeful
prompts; explains transformer mechanics and RLHF; recognizes
hallucination as structural. & Builds replicable prompting workflows;
teaches mechanics through guided discovery; publishes prompt
libraries. \\
Critical Evaluation & Accepts AI outputs at face value; rarely
cross-checks citations; treats fluent prose as accurate. & Triangulates
against scholarly databases; detects fabrications and bias routinely;
redesigns assessments. & Theorizes the source discovery--fabrication
paradox in disciplinary terms; mentors students into verification. \\
Human--AI Collaboration & Treats AI as prohibited or as an unsupervised
substitute; planning happens before or after AI, not with it. & Plans
lessons with AI as a contested co-participant; sequences
scaffold-removal assignments; models negotiation aloud. & Orchestrates
multi-tool workflows; designs assignments exploiting AI while building
independent competence. \\
Contextual Awareness & Imports activities from elite-institution
contexts without adaptation; unaware of access asymmetries. & Adapts
activities to learner background, language, and context; recognizes
differential affordance outcomes. & Designs equity-conscious AI
curricula at program or institutional level; contributes to MSI- or
region-specific adaptations. \\
Ethical Reasoning & Treats integrity as detection; relies on punitive
policy; cannot distinguish compliance from principled reasoning. & Uses
tier-differentiated citation pedagogy; distinguishes incidental,
substantive, and generative use. & Develops institutional AI governance
grounded in developmental ethics; treats first-time errors as
advancement opportunities. \\
Empowered Agency & Adopts AI because peers do; cannot articulate when to
restrain or refuse; reflection is episodic. & Maintains reflective
practice; models situated decision-making; recognizes principled non-use
as legitimate. & Sustains principled (Tier 3) reasoning under pressure;
participates in AI governance; cultivates civic critique in learners. \\
\bottomrule
\end{tabular}
\par\vspace{2pt}
\begin{minipage}{0.97\textwidth}\footnotesize \emph{Note.} Advancement in one pillar does not compensate for stagnation in another; advancement is recursive. MSI = minority-serving institution.\end{minipage}
\end{table*}
\subsection{Application Scenarios}\label{application-scenarios}

Two scenarios, summarized in Table 7, illustrate how the six pillars
and developmental rubric translate into concrete classroom decisions. Each is deliberately
specific: generic recommendations are the failure mode of prior
frameworks, and RAIL-Ed earns its theoretical claims by showing what
implementation looks like at the level of an actual unit.

\subsubsection{In-Service Teacher Context (K--12
Classroom)}\label{in-service-teacher-context-k12-classroom}

In a Grade 7 civics unit, RAIL-Ed shapes instruction at each phase.
Technical Fluency is built through a teacher-led prompt walkthrough
comparing how three different LLMs respond to the same civics question;
the comparison makes tokenization, training-data variation, and
probabilistic generation concrete. Critical Evaluation is practiced by
auditing an AI summary of a Bill of Rights primary source against the
original. Human--AI Collaboration is modeled by the teacher, who plans a
class debate aloud with the AI, narrating accept/override/ask-again
decisions. Contextual Awareness emerges when the class examines how the
AI handles county-level civic specifics and discusses whose knowledge is
centered in its training corpus. Ethical Reasoning is introduced through
a tier-graduated integrity conversation that asks what the school rule
says (Tier 1), what teachers and family expect (Tier 2), and what kind
of learner the student wants to be (Tier 3). Empowered Agency is
cultivated through a design-justice mini-project, supported by a
reflection journal: students propose one classroom rule for AI use and
argue for it to the school council. The unit operationalizes the K--12
ethical competencies that Tagare et al. (2026) and Sattelmaier and
Pawlowski (2025) identify as central to teaching with AI competencies
that Section 3 showed remain marginal in existing literacy frameworks.

\subsubsection{Pre-Service Teacher Context (Methods
Course)}\label{pre-service-teacher-context-methods-course}

In a pre-service teacher-education methods course, RAIL-Ed is built into
the semester's architecture. Technical Fluency is established through a
transformer-mechanics module paired with an iterative prompting task on
a primary source, with revisions logged. Critical Evaluation is
operationalized through a verification assignment in which pre-service
teachers locate three AI-suggested sources, confirm them in the library
database, and reflect on a fabrication they detected. Human--AI
Collaboration is enacted through a lesson-plan co-drafting exercise:
candidates generate a lesson plan with GenAI, revise it independently,
and submit both versions plus a reflection on what they changed and why.
Contextual Awareness is taught through a multilingual comparison
exercise examining whose voice AI editing preserves and whose it smooths
away (Warschauer et al., 2023). Ethical Reasoning is structured by a
course-wide AI citation protocol distinguishing incidental, substantive,
and generative use, with first-time disclosure errors treated as
developmental occasions. Empowered Agency is cultivated through an
end-of-term portfolio that includes a Reliance Negotiation Statement
(Hossain, 2026b), a reflective account in which candidates describe when
and why they used, restrained, or refused GenAI across the semester, and
a public-facing op-ed advocating an institutional AI policy position.

\begin{table*}[!tp]
\centering\small
\caption{RAIL-Ed Pillars Applied to In-Service and Pre-Service K--12 Teacher Contexts}
\begin{tabular}{@{}>{\RaggedRight\arraybackslash}p{\dimexpr 0.1960\textwidth-2\tabcolsep\relax}>{\RaggedRight\arraybackslash}p{\dimexpr 0.4036\textwidth-2\tabcolsep\relax}>{\RaggedRight\arraybackslash}p{\dimexpr 0.4003\textwidth-2\tabcolsep\relax}@{}}
\toprule
\textbf{Pillar} & \textbf{K--12 application (Grade 7 civics unit)} &
\textbf{Pre-service teacher application (methods course)} \\
\midrule
Technical Fluency & Teacher-led prompt walkthrough comparing how three
LLMs answer the same civics question; tokenization made concrete. &
Short transformer-mechanics module paired with iterative prompting on a
primary-source task; prompt revisions logged. \\
Critical Evaluation & Pupils audit AI summaries of a Bill of Rights
primary source against the original, identifying omissions and tone
shifts. & Verification assignment: candidates locate three AI-suggested
sources, confirm them in the library database, and reflect on a
fabrication they detected. \\
Human--AI Collaboration & Teacher plans a class debate aloud with AI,
narrating accept/override/ask-again decisions students will themselves
make. & Lesson-plan co-drafting: candidates generate a lesson plan with
GenAI, revise it independently, and submit both versions plus a
reflection on their changes. \\
Contextual Awareness & Class examines how AI handles county-level civic
specifics it is unlikely to represent well; discusses whose knowledge is
centered. & Candidates run a multilingual comparison of AI-edited drafts
in home language and English, examining whose voice AI preserves or
smooths away. \\
Ethical Reasoning & Tier-graduated integrity conversation: what the rule
says (Tier 1), what teachers and family expect (Tier 2), who you want to
be (Tier 3). & Course-wide citation protocol distinguishing incidental,
substantive, and generative use; first-time errors as developmental
occasions. \\
Empowered Agency & Reflection journal plus a design-justice
mini-project: students propose one classroom AI rule and argue for it to
the school council. & End-of-term portfolio with a Reliance Negotiation
Statement and a public-facing op-ed on institutional AI policy. \\
\bottomrule
\end{tabular}
\par\vspace{2pt}
\begin{minipage}{0.97\textwidth}\footnotesize \emph{Note.} Scenarios are illustrative implementations of the framework, not prescribed curricula; local adaptation is expected.\end{minipage}
\end{table*}
These scenarios demonstrate that RAIL-Ed accommodates the maturation of
each pillar across the K--12 teacher-preparation continuum, from
pre-service preparation through in-service practice. A teacher candidate
practicing Tier 1 and Tier 2 ethical reasoning in a methods course is
building the scaffolding from which Tier 3 principled reasoning matures
across years of classroom practice. The framework's integrative,
developmental, and dialectical commitments distinguish RAIL-Ed from the
additive competency lists the field has, to date, produced.

\section{Implications for Practice, Policy, and
Research}\label{implications-for-practice-policy-and-research}

The RAIL-Ed framework aims to reposition GenAI literacy as more than
functional knowledge or skills in using classroom tools. Earlier AI
literacy frameworks established important foundations for understanding,
using, and evaluating AI (Long \& Magerko, 2020; Ng et al., 2021), but
the rise of GenAI introduces new epistemic, pedagogical, and ethical
demands: teachers must now judge fluent but fallible outputs, negotiate
human-AI authorship, identify bias embedded across the AI life cycle,
and decide when AI supports or suppresses learning (Kasneci et al.,
2023; Lee et al., 2024). The significance of RAIL-Ed therefore lies in
its shift from competency acquisition to educational judgment. It treats
responsible GenAI literacy as an institutional and pedagogical project:
teacher preparation and training curricula must cultivate verification
and disciplinary reasoning, teacher learning must develop situated
professional judgment, policy must protect equity and accountability,
and research must validate evidence of actual learning.

\subsection{Implications for
Practice}\label{implications-for-practice}

The RAIL-Ed framework implies that GenAI literacy should reshape how
teacher preparation and training are designed. The central pedagogical
implication is that teachers need to shift from output-centered
instruction toward pedagogies that make reasoning, verification,
revision, and judgment visible. Teachers' responsible GenAI literacy is
treated as the intended outcome of this preparation, not simply their
technical fluency with a tool. Teacher candidates and in-service
teachers learn not only how to use GenAI, but also how to question a
source, check evidence, and notice when fluent language hides weak
reasoning or bias, so that they are equipped to model and scaffold this
judgment for the students they teach (Kasneci et al., 2023).

This shift requires embedding GenAI literacy within disciplinary
inquiry. For example, in science, teachers learn to examine how an AI
explanation simplifies causal mechanisms or misrepresents uncertainty,
and to design lessons that surface those simplifications for their
students. In social studies methods courses, teachers practice comparing
AI-generated claims with primary sources to detect hallucinations while
asking whose perspectives are presented or missed, then translate that
practice into classroom tasks. These pedagogical moves position GenAI,
for teachers, beyond a tool for productivity to an object of critique
and inquiry that they must first master before they can teach it.

The same shift appears in assessment, which now captures teachers' own
epistemic activity as GenAI users. A polished lesson plan, unit design,
or written reflection submitted in a methods course or
professional-development module no longer gives teacher educators the
same evidence it once did. A teacher candidate may have written it
independently, revised AI output carefully, or accepted generated text
with little understanding of the pedagogical reasoning behind it. The
final product alone may not tell the difference. For this reason,
teachers need evidence of how they arrived at a lesson, assessment, or
communication: what they accepted, rejected, verified, or questioned.
RAIL-Ed therefore treats responsible GenAI use as evidence that teachers
really develop this judgment instead of a rule they must obey.

Finally, pedagogical practice should cultivate discipline as well as use
in both teaching and learning. RAIL-Ed includes knowing when not to use
AI, when to slow down, when to seek human feedback, and when AI may
flatten uncertainty, voice, culture, or disciplinary complexity.
Teachers do not need to become computer scientists, but they do need a
basic understanding of how GenAI works. They should know, for example,
that LLMs generate likely sequences of words, not verified facts. This
helps explain why a model can produce a fluent answer, a false citation,
or a confident explanation that is only partly correct (Bender et al.,
2021). Without this understanding, teachers may trust AI outputs too
quickly. They may also reject technology entirely and miss situations
where it could support learning.

The practical implication of RAIL-Ed is that teachers should learn to
design learning environments with responsible AI use so that students
consistently practice disciplined decisions on AI's role in learning.

\subsection{Implications for Policy and
Governance}\label{implications-for-policy-and-governance}

RAIL-Ed suggests that school and institutional policy move beyond
reactive control of GenAI use toward governance that supports
responsible and authentic learning. Many current policies respond to
GenAI by focusing on plagiarism detection, disclosure rules, or bans on
AI-assisted work. These responses are understandable, with concerns
about authorship and academic integrity. However, they are insufficient
as a policy foundation because they treat GenAI primarily as a threat to
academic integrity. RAIL-Ed-informed stakeholders would ask whether
institutional rules help teachers develop judgment, accountability, and
agency in AI-assisted work.

Policy should therefore provide clearer guidance across different
professional situations. A teacher who uses AI to brainstorm lesson
ideas is not doing the same thing as a teacher who submits a fully
AI-generated lesson plan, assessment, or parent communication without
review. AI-assisted editing, resource curation, translation support, and
co-planning also raise different pedagogical and ethical questions.
Institutions need tiered guidance that names these differences and
connects them to professional purposes: when AI use is discouraged, when
it is permitted with disclosure, when it is scaffolded as part of
professional development, and when it is encouraged for critical
evaluation or accessibility. Otherwise, teacher-preparation programs and
school leaders may be left to guess what counts as acceptable use.

Governance must address due process and equity. Policy must account for
unequal access to paid AI tools, reliable devices, teacher support, and
dominant-language advantage. Without explicit equity provisions, GenAI
may widen existing educational inequalities. Finally, RAIL-Ed frames
policy as an institutional responsibility, not a classroom management
issue. Institutions should establish standards for privacy, data
protection, accessibility, teacher professional learning, and human
oversight. This aligns with broader international guidance emphasizing
human agency, inclusion, equity, privacy, transparency, and
accountability in AI governance (OECD, 2024; Miao \& Holmes, 2023).

\subsection{Implications for Future
Research}\label{implications-for-future-research}

RAIL-Ed also has implications for how GenAI literacy should be studied.
Instead of relying mainly on surveys or acceptance models, researchers
should track evidence of AI-assisted learning. This includes how
learners verify information, revise AI-generated text, explain their
choices, and decide when to use AI and when not. Longitudinal and
classroom-based studies would be especially useful because GenAI
literacy develops through repeated use, feedback, mistakes, and changing
expectations. Such evidence would allow researchers to distinguish
between superficial AI use and deeper forms of critical evaluation,
ethical reasoning, and empowered agency.

RAIL-Ed also encourages researchers to study GenAI literacy as a
situated educational practice. Teachers' responsible AI use depends on
preparation-program design, mentorship and coaching, institutional
policy, tool access, and disciplinary norms. Research should therefore
attend to the professional environment in which AI literacy is enacted.
Previous research has found that AI systems can intensify existing
inequities through unequal access, dominant-language advantage, and
biased outputs across the AI life cycle (Lee et al., 2024). The research
implication of RAIL-Ed is methodological as well as conceptual, and
studies should examine how human judgment develops within
socio-technical systems. RAIL-Ed suggests that the field needs stronger
alignment between constructs and evidence. When AI literacy is defined
as technical fluency, critical evaluation, human-AI collaboration,
contextual awareness, ethical reasoning, and empowered agency, research
instruments and analytic methods must capture these dimensions directly.
Otherwise, the field risks measuring convenience constructs, such as
frequency of use or perceived usefulness, while claiming to study
literacy. A RAIL-Ed-informed research agenda would therefore push AI
literacy scholarship toward richer evidence of teacher learning and
stronger construct validity where responsible AI judgment becomes
possible.

\section{Limitations and Directions for Future Work}\label{limitations-and-directions-for-future-work}

\subsection{Conceptual Scope and
Boundaries}\label{conceptual-scope-and-boundaries}

The RAIL-Ed framework is conceptually oriented and necessarily bounded
in scope. It synthesizes critical pedagogy, pragmatist inquiry,
sociocultural theory, and human-centered AI design, yet it does not
exhaust the theoretical resources available for thinking about GenAI in
education (Long \& Magerko, 2020; Ng et al., 2021). Adjacent traditions,
including posthumanist and feminist epistemologies, decolonial
perspectives, and disability studies in education, could extend the
framework in productive ways and surface tensions that the present
synthesis addresses only partially (Nemorin, 2024). The six pillars
operate at a level of abstraction that requires interpretive translation
into specific disciplines, grade levels, and institutional cultures (Ng
et al., 2021; Zhang et al., 2025). The framework does not prescribe
particular instructional methods or assessment instruments, nor does it
claim universal applicability across educational settings. It offers,
rather, a coherent scaffold that scholars and practitioners can adapt,
contest, and refine in light of context-specific commitments and
constraints.

\subsection{Areas for Empirical
Research}\label{areas-for-empirical-research}

A further limitation concerns the framework\textquotesingle s
evidentiary base. Several of the constructs that inform RAIL-Ed,
including the three-tier ethical reasoning model developed in Section
4.3 and the reliance-negotiation account behind its integrity pedagogy,
derive from the lead author\textquotesingle s recent mixed-methods
research (Hossain, 2026a, 2026b), conducted at a single minority-serving
institution and not yet independently replicated or peer reviewed.
Independent testing across institutions, populations, and grade levels
is a precondition for the framework\textquotesingle s broader claims,
and we flag this dependency so adopters can weigh its propositions
accordingly.

This need is not unique to RAIL-Ed; recent umbrella reviews show that
empirical work on AI literacy remains uneven across grade levels,
methodological traditions, and disciplinary contexts (Fu et al., 2025).
Two directions follow specifically from the framework\textquotesingle s
internal structure, complementing the broader agenda in Section 5.3.
First, studies might examine relationships among the pillars, such as
how technical fluency mediates ethical reasoning or how contextual
awareness shapes human-AI collaboration (Lee et al., 2024). Second,
design-based research could refine curricular and instructional
approaches grounded in these interdependencies, clarifying which pillars
require domain-specific elaboration and which travel across content
areas.

\subsection{Potential for Cross-National
Adaptation}\label{potential-for-cross-national-adaptation}

Although the framework draws on broadly applicable theoretical
traditions, its instantiation will vary considerably across national,
cultural, and policy contexts. Educational systems differ in their
governance structures, infrastructural capacities, linguistic ecologies,
and prevailing pedagogical philosophies, and these differences shape
what responsible AI literacy can and should mean in particular settings
(OECD, 2024; Miao \& Holmes, 2023). The framework therefore invites
cross-national adaptation. Scholars working in the Global South have
noted that frameworks developed primarily in North American and European
contexts often underrepresent local epistemologies, languages, and
infrastructural realities, and that the ethics of AI in education itself
reflects Eurocentric assumptions that warrant decolonial interrogation
(Nemorin, 2024). Comparative research that situates the RAIL-Ed
framework within emerging regional policy environments, including the
African Union\textquotesingle s (2024) Continental AI Strategy and
parallel initiatives across ASEAN and Latin American educational
ministries, could surface productive tensions and reveal opportunities
for context-sensitive refinement. Such work would strengthen the
framework\textquotesingle s global relevance and its capacity to support
equitable, culturally grounded responses to GenAI in education.

\section{Conclusion}\label{conclusion}

This paper has proposed the RAIL-Ed framework as a theoretically
integrated response to the conceptual fragmentation in current GenAI
literacy discourse (Long \& Magerko, 2020; Ng et al., 2021). The
framework articulates six interdependent pillars: technical fluency,
critical evaluation, human-AI collaboration, contextual awareness,
ethical reasoning, and empowered agency. Drawing on
Freire\textquotesingle s (1970/2000) pedagogy of critical consciousness,
Dewey\textquotesingle s (1938) reflective inquiry,
Vygotsky\textquotesingle s (1978) sociocultural theory of mediated
cognition, and Shneiderman\textquotesingle s (2020) principles of
human-centered AI, RAIL-Ed positions GenAI literacy as a relational,
situated, and ethically grounded practice. The framework offers a
coherent scaffold for curriculum design, teacher education, professional
development, and policy formation across K--12 contexts (OECD, 2024;
Miao \& Holmes, 2023).

Central to this position is RAIL-Ed\textquotesingle s rejection of
deficit-oriented approaches, which attribute educational challenges to
perceived inadequacies in learners or communities rather than to the
structural conditions shaping access and participation (Davis \& Museus,
2019; Tewell, 2020). In GenAI contexts, deficit framing recasts
students\textquotesingle{} struggles as insufficient digital readiness
and narrows literacy to prompt compliance, obscuring the algorithmic and
institutional conditions that produce those struggles (Kay et al., 2024;
Lee et al., 2024). Freire\textquotesingle s (1970/2000) critique of the
banking model offers a corrective: learners are not vessels to be filled
with AI skills but active participants capable of interrogating the
systems that mediate their knowledge.

The responsible integration of GenAI in education is not a technical
problem to be solved but an ongoing pedagogical, ethical, and political
project.

Researchers should test, contest, and extend the framework through
studies attentive to context, culture, and equity, particularly in
settings underrepresented in the literature (Nemorin, 2024; Zhang et
al., 2025). Educators can treat it not as a prescription but as a
generative scaffold for designing experiences in which students question
GenAI, collaborate with it critically, and shape its place in their
learning (Selwyn, 2022). Policymakers, in turn, should align curricular,
infrastructural, and regulatory commitments with the principles of human
oversight, equity, and justice articulated in international and regional
guidance (African Union, 2024; OECD, 2024; Miao \& Holmes, 2023).
Building a responsible GenAI literacy is necessarily collective,
requiring sustained collaboration across disciplines, sectors, and
geographies and a willingness to reframe AI literacy as a shared
commitment to human flourishing in an algorithmic age.

\FloatBarrier
\section*{References}
\begingroup
\setlength{\parindent}{-0.2in}
\setlength{\leftskip}{0.2in}
\setlength{\parskip}{2.5pt}

African Union. (2024). \emph{Continental artificial intelligence
strategy: Harnessing AI for Africa\textquotesingle s development and
prosperity.}
\url{https://au.int/sites/default/files/documents/44004-doc-EN-_Continental_AI_Strategy_July_2024.pdf}

Allen, L. K., \& Kendeou, P. (2024). ED-AI Lit: An interdisciplinary
framework for AI literacy in education. \emph{Policy Insights from the
Behavioral and Brain Sciences, 11}(1), 3--10.
\url{https://doi.org/10.1177/23727322231220339}

Annapureddy, R., Fornaroli, A., \& Gatica-Perez, D. (2025). Generative
AI literacy: Twelve defining competencies. \emph{Digital Government:
Research and Practice, 6}(1), Article 13, 1--21.
\url{https://doi.org/10.1145/3685680}

Arrington, C., McVey, D., Mativo, J., \& Pidaparti, R. M. (2025).
Enhancing teacher effectiveness with AI-based prompt engineering: A
proof of concept. \emph{Journal of STEM Education: Innovations and
Research, 26}(2), 5--10.
\url{https://doi.org/10.63504/jstem.v26i2.2711}

Bae, H., Hur, J., Park, J., Choi, G. W., \& Moon, J. (2024). Pre-service
teachers\textquotesingle{} dual perspectives on generative AI: Benefits,
challenges, and integrating into teaching and learning. \emph{Online
Learning, 28}(3), 131--156.
\url{https://doi.org/10.24059/olj.v28i3.4543}

Barbieri, W., \& Nguyen, N. (2025). Generative AI as a "placement
buddy": Supporting pre-service teachers in work-integrated learning,
self-management and crisis resolution. \emph{Australasian Journal of
Educational Technology, 41}(2), 34--49.
\url{https://doi.org/10.14742/ajet.10035}

Bender, E. M., Gebru, T., McMillan-Major, A., \& Shmitchell, S. (2021).
On the dangers of stochastic parrots: Can language models be too big? In
\emph{Proceedings of the 2021 ACM Conference on Fairness,
Accountability, and Transparency} (pp. 610--623). Association for
Computing Machinery.
\url{https://doi.org/10.1145/3442188.3445922}

Bower, M., Torrington, J., Lai, J. W. M., Petocz, P., \& Alfano, M.
(2024). How should we change teaching and assessment in response to
increasingly powerful generative artificial intelligence? Outcomes of
the ChatGPT teacher survey. \emph{Education and Information
Technologies, 29}(12), 15403--15439.
\url{https://doi.org/10.1007/s10639-023-12405-0}

Bukar, U. A., Sayeed, M. S., Razak, S. F. A., Yogarayan, S., \& Sneesl,
R. (2024). Prioritizing ethical conundrums in the utilization of ChatGPT
in education through an analytical hierarchical approach.
\emph{Education Sciences, 14}(9), Article 959.
\url{https://doi.org/10.3390/educsci14090959}

Cheah, Y. H., Lu, J., \& Kim, J. (2025). Integrating generative
artificial intelligence in K--12 education: Examining
teachers\textquotesingle{} preparedness, practices, and barriers.
\emph{Computers and Education: Artificial Intelligence, 8,} Article
100363.
\url{https://doi.org/10.1016/j.caeai.2025.100363}

Chiu, T. K. F., Ahmad, Z., Ismailov, M., \& Sanusi, I. T. (2024). What
are artificial intelligence literacy and competency? A comprehensive
framework to support them. \emph{Computers and Education Open, 6,}
Article 100171.
\url{https://doi.org/10.1016/j.caeo.2024.100171}

Choi, Y. S. (2025). Earth science simulations with generative artificial
intelligence (GenAI). \emph{Journal of University Teaching and Learning
Practice, 22}(1).

\url{https://doi.org/10.53761/nf1yqr46}

Davis, L. P., \& Museus, S. D. (2019). What is deficit thinking? An
analysis of conceptualizations of deficit thinking and implications for
scholarly research. \emph{Currents: Journal of Diversity Scholarship for
Social Change, 1}(1).
\url{https://doi.org/10.3998/currents.17387731.0001.110}

Dewey, J. (1938). \emph{Experience and education.} Macmillan.

European Union. (2024). REGULATION (EU) 2024/1689 OF THE EUROPEAN
PARLIAMENT AND OF THE COUNCIL of 13 June 2024 laying down harmonised
rules on artificial intelligence (Artificial Intelligence Act). In
\emph{eur-lex.europa.eu}. Official Journal of the European Union.
\url{https://eur-lex.europa.eu/eli/reg/2024/1689/oj}

Feldman-Maggor, Y., Blonder, R., \& Alexandron, G. (2025). Perspectives
of generative AI in chemistry education within the TPACK framework.
\emph{Journal of Science Education and Technology, 34}(1), 1--12.
\url{https://doi.org/10.1007/s10956-024-10147-3}

Freire, P. (2000). \emph{Pedagogy of the oppressed} (M. B. Ramos,
Trans.; 30th anniversary ed.). Continuum. (Original work published 1970)

Fu, Y., Weng, Z., \& Wang, J. (2025). Examining AI use in educational
contexts: A scoping meta-review and bibliometric analysis.
\emph{International Journal of Artificial Intelligence in Education,
35}(3), 1388--1444.
\url{https://doi.org/10.1007/s40593-024-00442-w}

Gibson, J. J. (1979). \emph{The ecological approach to visual
perception.} Houghton Mifflin.

Gruenhagen, J. H., Sinclair, P. M., Carroll, J.-A., Baker, P. R. A.,
Wilson, A., \& Demant, D. (2024). The rapid rise of generative AI and
its implications for academic integrity: Students\textquotesingle{}
perceptions and use of chatbots for assistance with assessments.
\emph{Computers and Education: Artificial Intelligence, 7,} Article
100273.
\url{https://doi.org/10.1016/j.caeai.2024.100273}

Hossain, S. (2026a). \emph{Measuring college students' reliance on large
language models in academic writing: A sequential explanatory
mixed-methods study at a minority-serving institution} (Doctoral
dissertation, University of Maryland, Baltimore County).

Hossain, S. (2026b). The reliance negotiation framework: A dynamic
process model of student LLM engagement in academic writing.
\emph{arXiv preprint} arXiv:2604.16772.

Jin, Y., Martinez-Maldonado, R., Gašević, D., \& Yan, L. (2025). GLAT:
The generative AI literacy assessment test. \emph{Computers and
Education: Artificial Intelligence, 9}, Article 100436.
\url{https://doi.org/10.1016/j.caeai.2025.100436}

Jorolan, J., Cabillo, F., Batucan, R. R., Camansi, C. M., Etoquilla, A.,
Gapo, J., Hallarte, D. K., Milano, M. L., Gonzales, R., \& Gonzales, G.
(2025). Development and validation of moral ascendancy and dependency in
AI integration (MAD-AI) scale for teachers. \emph{Computers \&
Education, 235,} Article 105346.
\url{https://doi.org/10.1016/j.compedu.2025.105346}

Joseph, S. (2023). Large language model-based tools in language teaching
to develop critical thinking and sustainable cognitive structures.
\emph{\href{https://doi.org/10.1016/j.compedu.2025.105346}{\hl{Rupkatha
Journal}},}
\emph{15}\href{https://doi.org/10.21659/rupkatha.v15n4.13}{(4).
\ul{https://doi.org/10.21659/rupkatha.v15n4.13}}

Kasneci, E., Sessler, K., Küchemann, S., Bannert, M., Dementieva, D.,
Fischer, F., Gasser, U., Groh, G., Günnemann, S., Hüllermeier, E.,
Krusche, S., Kutyniok, G., Michaeli, T., Nerdel, C., Pfeffer, J.,
Poquet, O., Sailer, M., Schmidt, A., Seidel, T., \ldots{} Kasneci, G.
(2023). ChatGPT for good? On opportunities and challenges of large
language models for education. \emph{Learning and Individual
Differences, 103,} 102274.
\url{https://doi.org/10.1016/j.lindif.2023.102274}

Kay, J., Kasirzadeh, A., \& Mohamed, S. (2024). Epistemic injustice in
generative AI. \emph{Proceedings of the AAAI/ACM Conference on AI,
Ethics, and Society, 7}(1), 684--697.
\url{https://doi.org/10.1609/aies.v7i1.31671}

Kohlberg, L. (1984). \emph{The psychology of moral development: The
nature and validity of moral stages}. Harper \& Row.

Kong, S. C., Yang, Y., \& Hou, C. (2024). Examining teachers'
behavioural intention of using generative artificial intelligence tools
for teaching and learning based on the extended technology acceptance
model. \emph{Computers and Education: Artificial Intelligence},
\emph{7}, Article 100328.
\url{https://doi.org/10.1016/j.caeai.2024.100328}

Laine, J., Minkkinen, M., \& Mäntymäki, M. (2025). Understanding the
ethics of generative AI: Established and new ethical principles.
\emph{Communications of the Association for Information Systems},
\emph{56}, 1--25.
\url{https://doi.org/10.17705/1CAIS.05601}

Lee, J., Hicke, Y., Yu, R., Brooks, C., \& Kizilcec, R. F. (2024). The
life cycle of large language models in education: A framework for
understanding sources of bias. \emph{British Journal of Educational
Technology, 55}(5), 1982--2002.
\url{https://doi.org/10.1111/bjet.13505}

Long, D., \& Magerko, B. (2020). What is AI literacy? Competencies and
design considerations. In \emph{Proceedings of the 2020 CHI Conference
on Human Factors in Computing Systems}, 1--16. Association for Computing
Machinery.
\url{https://doi.org/10.1145/3313831.3376727}

MacDowell, P., Moskalyk, K., Korchinski, K., \& Morrison, D. (2024).
Preparing educators to teach and create with generative artificial
intelligence. \emph{Canadian Journal of Learning and Technology, 50}(4),
1--23.
\url{https://doi.org/10.21432/cjlt28606}

Memarian, B., \& Doleck, T. (2024). Human-in-the-loop in artificial
intelligence in education: A review and entity-relationship (ER)
analysis. \emph{Computers in Human Behavior Artificial Humans},
\emph{2}(1), Article 100053.
\url{https://doi.org/10.1016/j.chbah.2024.100053}

Miao, F., \& Cukurova, M. (2024). AI
competency framework for teachers. UNESCO.
\url{https://doi.org/10.54675/ZJTE2084}

Miao, F., \& Holmes, W. (2023). \emph{Guidance for Generative AI in
Education and Research}. UNESCO Publishing.
\url{https://unesdoc.unesco.org/ark:/48223/pf0000386693}

Miao, F., Holmes, W., Huang, R., \& Zhang, H. (2021). \emph{AI and
education: Guidance for policy-makers.} UNESCO.
\url{https://doi.org/10.54675/pcsp7350}

Miao, F., Shiohira, K., \& Lao, N. (2024). \emph{AI competency framework
for students}. UNESCO.
\url{https://doi.org/10.54675/JKJB9835}

Mills, K., Ruiz, P., Lee, K., Coenraad, M., Fusco, J., Roschelle, J., \&
Weisgrau, J. (2024). \emph{AI literacy: A framework to understand,
evaluate, and use emerging technology.} Digital Promise.
https://doi.org/10.51388/20.500.12265/218

Mollick, E. R., \& Mollick, L. (2023). \emph{Assigning AI: Seven
approaches for students, with prompts} (Preprint). arXiv.
\url{https://doi.org/10.48550/arXiv.2306.10052}

Nemorin, S. (2024). Towards decolonising
the ethics of AI in education. Globalisation, Societies and Education,
22(1), 1260--1272.
\url{https://doi.org/10.1080/14767724.2024.2333821}

Ng, D. T. K., Leung, J. K. L., Chu, S. K. W., \& Qiao, M. S. (2021).
Conceptualizing AI literacy: An exploratory review. \emph{Computers and
Education: Artificial Intelligence, 2,} Article 100041.
\url{https://doi.org/10.1016/j.caeai.2021.100041}

Ng, D. T. K., Chan, E. K. C., \& Lo, C. K. (2025). Opportunities,
challenges and school strategies for integrating generative AI in
education. \emph{Computers and Education: Artificial Intelligence, 8,}
Article 100373.
\url{https://doi.org/10.1016/j.caeai.2025.100373}

OECD. (2019). \emph{OECD future of education and skills 2030: OECD
learning compass 2030. A series of concept notes.} OECD Publishing.
\url{https://www.oecd.org/content/dam/oecd/en/about/projects/edu/education-2040/1-1-learning-compass/OECD_Learning_Compass_2030_Concept_Note_Series.pdf}

OECD. (2024). The OECD Artificial Intelligence (AI) Principles. In
\emph{oecd.ai}. OECD.
\url{https://oecd.ai/en/ai-principles}

OECD/European Commission. (2026). \emph{Empowering learners for the age
of AI: An AI literacy framework for primary and secondary education}.
OECD Publishing.
\url{https://doi.org/10.1787/65cd27d4-en}

Ofosu-Asare, Y. (2025). Guiding the future: Developing an ethical
framework for generative AI use in education. \emph{International
Journal of Information and Learning Technology, 43}(1), 26 -- 44.
\url{https://doi.org/10.1108/IJILT-06-2024-0113}

Park, J. (2025). A systematic literature review of generative artificial
intelligence (GenAI) literacy in schools. \emph{Computers and Education:
Artificial Intelligence, 9,} Article 100487.
\url{https://doi.org/10.1016/j.caeai.2025.100487}

Roe, J., Perkins, M., \& Furze, L. (2025). Reflecting reality,
amplifying bias? Using metaphors to teach critical AI literacy.
\emph{Journal of Interactive Media in Education, 2025}(1), Article 18.
\url{https://doi.org/10.5334/jime.961}

Sattelmaier, L., \& Pawlowski, J. (2025). Be AIware! An AI competency
model for K--12 education. \emph{Social Sciences \& Humanities Open,
12,} Article 101838.
\url{https://doi.org/10.1016/j.ssaho.2025.101838}

Selwyn, N. (2022). The future of AI and education: Some cautionary
notes. \emph{European Journal of Education, 57}(4), 620--631.
\url{https://doi.org/10.1111/ejed.12532}

Shneiderman, B. (2020). Human-centered artificial intelligence:
Reliable, safe \& trustworthy. \emph{International Journal of
Human--Computer Interaction, 36}(6), 495--504.
\url{https://doi.org/10.1080/10447318.2020.1741118}

Shulman, L. S. (1986). Those who understand: Knowledge growth in
teaching. \emph{Educational Researcher, 15}(2), 4--14.
\url{https://doi.org/10.3102/0013189X015002004}

Şimşek, N. (2025). Integration of ChatGPT in mathematical story-focused
5E lesson planning: Teachers and pre-service teachers\textquotesingle{}
interactions with ChatGPT. \emph{Education and Information Technologies,
30}, 11391--11462.
\url{https://doi.org/10.1007/s10639-024-13258-x}

Sourati, Z., Ziabari, A. S., \& Dehghani, M. (2026). The homogenizing
effect of large language models on human expression and thought.
\emph{Trends in Cognitive Sciences.} Advance online publication.
\url{https://doi.org/10.1016/j.tics.2026.01.003}

Sperling, K., Stenberg, C. J., McGrath, C., Åkerfeldt, A., Heintz, F.,
\& Stenliden, L. (2024). In search of artificial intelligence (AI)
literacy in teacher education: A scoping review. \emph{Computers and
Education Open, 6,} Article 100169.
\url{https://doi.org/10.1016/j.caeo.2024.100169}

Su, J., \& Yang, W. (2023). Unlocking the power of ChatGPT: A framework
for applying generative AI in education. \emph{ECNU Review of Education,
6}(3), 355--366.
\url{https://doi.org/10.1177/20965311231168423}

Tagare, D., Karki, T., \& Yu, W. (2025). K--12
teachers\textquotesingle{} ethical competencies for AI literacy:
Insights from a systematic literature review. \emph{Computers \&
Education, 239,} Article 105435.
\url{https://doi.org/10.1016/j.compedu.2025.105435}

Tan, Q. (2025). Reimagining teacher development in the era of generative
AI: A scoping review. \emph{Teaching and Teacher Education}, 168,
Article 105236.
\url{https://doi.org/10.1016/j.tate.2025.105236}

Tate, T. P., Ritchie, D. R., \& Warschauer, M. (2026). \emph{Generative
AI as a mediational agent: Rethinking learning in sociocultural theory}
(Preprint). EdArXiv.
\url{https://doi.org/10.35542/osf.io/wcpj5_v1}

Tewell, E. (2020). The problem with grit: Dismantling deficit thinking
in library instruction. \emph{portal: Libraries and the Academy, 20}(1),
137--159.
\url{https://doi.org/10.1353/pla.2020.0007}

Thomas, J., \& Harden, A. (2008). Methods for the thematic synthesis of
qualitative research in systematic reviews. \emph{BMC Medical Research
Methodology, 8,} Article 45.
\url{https://doi.org/10.1186/1471-2288-8-45}

Touretzky, D., Gardner-McCune, C., Martin, F., \& Seehorn, D. (2019).
Envisioning AI for K--12: What should every child know about AI?
\emph{Proceedings of the AAAI Conference on Artificial Intelligence,
33}(1), 9795--9799.
\url{https://doi.org/10.1609/aaai.v33i01.33019795}

Vygotsky, L. S. (1978). \emph{Mind in society: The development of higher
psychological processes} (M. Cole, V. John-Steiner, S. Scribner, \& E.
Souberman, Eds.). Harvard University Press.

Walker, L. O., \& Avant, K. C. (2005). \emph{Strategies for theory
construction in nursing} (4th ed.). Pearson Prentice Hall.

Wang, Y., Zhang, T., Yao, L., \& Seedhouse, P. (2025). A scoping review
of empirical studies on generative artificial intelligence in language
education. \emph{Innovation in Language Learning and Teaching, 1-28}.
\url{https://doi.org/10.1080/17501229.2025.2509759}

Warschauer, M., Tseng, W., Yim, S., Webster, T., Jacob, S., Du, Q., \&
Tate, T. (2023). The affordances and contradictions of AI-generated text
for writers of English as a second or foreign language. \emph{Journal of
Second Language Writing, 62,} Article 101071.
\url{https://doi.org/10.1016/j.jslw.2023.101071}

Yan, W., Liu, Y.-l., Mamaeva, V., Dong, F., Tao, G., Li, R., \& Yang, H.
(2026). Generative AI literacy: Scale development and its influence on
privacy protection behaviors and information verification behaviors.
\emph{Telecommunications Policy, 50}(2), Article 103117.
\url{https://doi.org/10.1016/j.telpol.2025.103117}

Yang, S., \& Appleget, C. (2025). Besties with bots: GenAI as a lesson
plan evaluator for preservice teachers. \emph{Journal of Digital
Learning in Teacher Education}, \emph{41}(2--3), 101--115.
\url{https://doi.org/10.1080/21532974.2025.2474932}

Yim, I. H. Y., \& Su, J. (2025). Artificial intelligence literacy
education in primary schools: A review. \emph{International Journal of
Technology and Design Education}, \emph{35}, 2175--2204.
\url{https://doi.org/10.1007/s10798-025-09979-w}

\hl{Yu, X., \& Pian, Y.} (2025). Towards actionable GenAI-classroom
integration: A social role framework for teacher's practice. In A. I.
Cristea, E. Walker, Y. Lu, O. C. Santos, \& S. Isotani (Eds.),
\emph{Artificial intelligence in education. Posters and late breaking
results, workshops and tutorials, industry and innovation tracks,
practitioners, doctoral consortium, blue sky, and wideAIED. AIED 2025}
(Vol. 2590, pp. 158--167). Springer.
\url{https://doi.org/10.1007/978-3-031-99261-2_15}

Zhang, Y., Lai, C., \& Gu, M. M. Y. (2025). Becoming a teacher in the
era of AI: A multiple-case study of pre-service teachers' investment in
AI-facilitated learning-to-teach practices. \emph{System, 133}, Article
103746.
\url{https://doi.org/10.1016/j.system.2025.103746}

Zhou, C., Ahmadi, S., Li, L., \& Hossain, S. (in press). Conceptualizing
generative AI literacy in teacher education: A systematic review.

\endgroup

\section*{Author Contributions}

\textbf{Shahin Hossain:} Conceptualization, Methodology, Investigation and Validation, Writing -- original draft (Section 4), Writing -- review \& editing, Proofreading.
\textbf{Sima Ahmadi:} Investigation (literature review), Writing -- original draft (Section 3, with contributions to other sections), Writing -- review \& editing.
\textbf{Leqi Li:} Conceptualization, Methodology, Investigation (literature review), Writing -- original draft (contributions to Section 3), Writing -- review \& editing.
\textbf{Idowu David Awoyemi:} Conceptualization, Methodology, Investigation and Validation, Writing -- original draft (Sections 1, 6, and 7), Writing -- review \& editing.
\textbf{Wei Huang:} Conceptualization, Writing -- original draft (Section 5), Reference curation and in-text citation verification, Writing -- review \& editing.
\textbf{Chenxi Zhou:} Investigation (literature review), Writing -- original draft (contributions to Section 3 and other sections), Writing -- review \& editing, Reference and in-text citation verification.
\textbf{Jujia Li:} Conceptualization, Writing -- original draft (Section 5), Writing -- review \& editing.
\textbf{Samaa Haniya:} Writing -- review \& editing, Validation, Expert Feedback.
\textbf{Shapla Khanam:} Writing -- review \& editing, Validation.
\textbf{Tasbirun Mashreka Subaha:} Reference curation and in-text citation verification.
All authors contributed to critical revision of the manuscript and read and approved the final version. All authors agree to be accountable for all aspects of the work in ensuring that questions related to the accuracy or integrity of any part of the work are appropriately investigated and resolved.

\section*{Declaration of generative AI and AI-assisted technologies in the writing process}

During the preparation of this work, the authors used Claude (Anthropic;
Fable 5, Opus 4.8, and Sonnet 4.6) and ChatGPT (OpenAI; GPT-5.5) to
improve language clarity and style; copyedit the manuscript; check and
correct reference formatting and bibliographic details; verify internal
consistency of citations, terminology, and cross-references; format
tables and prepare figures; and check compliance with journal formatting
requirements. All conceptual framing, arguments, analyses, and
interpretations are the authors\textquotesingle{} own. After using these
tools, the authors reviewed, verified, and edited all content and took
full responsibility for the content of the published article.

\balance
\end{document}